\documentclass[jou,floatsintext]{apa6}

\usepackage[american]{babel}
\usepackage[utf8]{inputenc}
\usepackage[T1]{fontenc}
\usepackage{csquotes}
\usepackage[backend=biber,useprefix=true,style=apa,url=false,doi=false,eprint=false]{biblatex}
\usepackage{graphicx}
\usepackage{amsmath}
\usepackage{mathptmx}      
\usepackage{amsfonts}
\graphicspath{{figs/}{../figs/}}
\usepackage{blindtext}
\usepackage{todonotes}
\usepackage{subfiles}

\DeclareLanguageMapping{american}{american-apa}
\let \citet \textcite
\let \citep \parencite

\title{Adaptive nonparametric psychophysics}
\shorttitle{Adaptive nonparametric psychophysics}

\sixauthors{Lucy Owen}{Jonathan Browder}{Benjamin Letham}{Gideon Stocek}{Chase Tymms}{Michael Shvartsman}
\sixaffiliations{Dartmouth College\footnote{work done while on internship at Facebook Reality Labs}}{Facebook Reality Labs}{Facebook}{Facebook Reality Labs}{Facebook Reality Labs}{Facebook Reality Labs}

\leftheader{Owen}


\abstract{We introduce a new set of models and adaptive psychometric testing methods for multidimensional psychophysics. In contrast to traditional adaptive staircase methods like PEST and QUEST, the method is multi-dimensional and does not require a grid over contextual dimensions, retaining sub-exponential scaling in the number of stimulus dimensions. In contrast to more recent multi-dimensional adaptive methods, our underlying model does not require a parametric assumption about the interaction between intensity and the additional dimensions. In addition, we introduce a new active sampling policy that explicitly targets psychometric detection threshold estimation and does so substantially faster than policies that attempt to estimate the full psychometric function (though it still provides estimates of the function, albeit with lower accuracy). Finally, we introduce \texttt{AEPsych}, a user-friendly open-source package for nonparametric psychophysics that makes these technically-challenging methods accessible to the broader community.}

\begin{document}
\maketitle

\section{Introduction}

An objective of psychophysicial research is to estimate the transformation of the perceptual system on a stimulus, and to understand how this transformation operates across contexts and different stimuli. This transformation function cannot be measured directly, but is rather inferred from subjective participant responses. Conventional psychophysical experiments do this by averaging participant responses to each stimulus over many repetitions to arrive at average detection probabilities, which are then modeled using some parametric form (often linear). This requires large number of trials per point in the input domain of the psychometric function (e.g.\ stimulus intensity). Furthermore, classical methods attempt to densely sample this input domain, suffering from the \emph{curse of dimensionality} wherein the number of points needed to fill a space grows exponentially in the number of dimensions. Additional challenges arise when extending models to include non-linear interactions between stimulus features (i.e.\ when response probabilities have a complex relationship to multiple stimulus parameters) or when the stimulus features are noisy or coded on the wrong scale (i.e.\ stimulus features must be coded in the correct units for classical models like the Weber-Fechner law to apply). Finally, traditional methods need to perform different experiments to estimate different aspects of the psychometric transfer function such as detection and discrimination thresholds. This is required because the standard model for discrimination, Weber's law, breaks down at low intensities, so so-called \emph{sub-threshold} and \emph{supra-threshold} behavior are modeled separately \citep[e.g.][]{Georgeson1975,Guan2016}.

Our work, while not first in addressing some of these challenges, is among the first in addressing them all together. First, we treat the value of the psychometric transfer function as a latent variable, rather than computing it from averages post-hoc. This is consistent with common practice in psychology and some psychometric fitting toolboxes \citep[e.g.]{Schutt2016}, though less common in experimental usage. Practically, it means we almost never need to sample the exact same point twice, since we can learn more from a closely adjacent point. Second, we sample the input domain adaptively based on participant responses rather than using a predetermined set of stimuli, which allows us to partially mitigate the curse of dimensionality. In particular, we introduce \emph{level set estimation} (LSE), an objective that explicitly targets estimation of psychometric thresholds. Third, we use a nonparametric model for the psychometric transfer function $f$, which allows us to make fewer assumptions about the shape of the psychometric curve. This means we can handle nonlinear interactions between input features, and nonlinear input scaling. And finally, our model jointly covers sub-threshold and supra-threshold behavior in a formal generalization of both detection and discrimination models, and as such allows for estimation of both slopes and intercepts of the psychometric function from a single experiment.

To enable practitioners to apply our method to their own domains, we are making available a public implementation of all the methods in this paper, as well as the evaluation and benchmark code used to generate our results. This code can be used to evaluate new models and test functions, and additionally supports interfacing with Python, Matlab, or Unity for human-in-the-loop experimentation.

To demonstrate the distinct benefits of both our novel modeling and stimulus selection contributions, we show extensive simulation results using both previously-reported and novel test functions derived from real data. Specifically, we demonstrate sample efficiency benefits of up to 10x even relative to previous adaptive methods, and potentially far more relative to the method of constant stimuli, without the strong parametric assumptions of past adaptive methods.

The paper proceeds as follows. First, we provide a longer background review of psychophysics theory and adaptive methods for psychophysics. Next, we show how a particular Bayesian nonparametric model, the Gaussian process (GP) with a probit likelihood, can be thought of as a formal generalization of classical theory, supporting its use for adaptive data collection on theoretical as well as empirical grounds. We then demonstrate the benefits of the method with a thorough empirical study comparing to adaptive psychophysics methods in common use. Finally, we provide an overview of the key features of \texttt{AEPsych}, our new package for adaptive experimentation in psychophysics, and conclude with a discussion and broader outlook.

\section{Background}

\subsection{Classical psychophysical methods}
One of the experimental objectives of classical psychophysics is to measure three quantities of interest: the
detection threshold (DT),  the just-noticeable-difference (JND), and the point of subjective equality (PSE).
The DT is defined as the lowest stimulus intensity at which the observer will correctly detect a stimulus with
some average probability. The JND is defined as the difference in intensities between two stimuli such that
the observer will correctly detect the difference with some average probability, often
taken to be 0.75. The PSE is defined as the stimulus intensity where two stimuli appear equal, i.e.\ a JND
for probability 0.5. These last two quantities are also sufficient statistics for the parameters of the full psychometric function under classical assumptions, since they essentially specify the slope and intercept of a linear model.

Standard methods of classical psychophysics are the method of constant stimuli, the method of limits, and the method of adjustment. The first of these is a standard randomized experiment where participants are shown repeated stimuli from a predefined set (often a fixed grid over the stimulus domain), and in the second, stimuli are shown in ascending or descending order. In both cases, participants are asked to respond as to whether they detected the stimulus or difference. In the method of adjustment, participants are not asked to respond to stimuli, rather they are asked to adjust a second target stimulus until it matches a predetermined probe.

A deep discussion of the relative advantages and disadvantages of classical methods is beyond the scope of the present work (though see \citet{Klein2001}), but none of these methods are suitable for evaluating stimuli with more than one or two dimensions. For a conventional method of constant stimuli grid, the number of stimuli grows exponentially in both the number of dimensions and the number of points per dimension, yielding experiments that take upwards of 5 hours per observer \citep[e.g.][]{Guan2016,Wier1977}, though sparser irregular grids are theoretically possible. For the method of limits, the ordering is typically in just one dimension, requiring a grid over the other dimensions and thus suffering from the same issue. A secondary concern with these methods is that the same exact stimulus must be repeated many times in order to estimate its response probability, and no information is shared across similar stimuli. For the method of adjustment, the search of the stimulus space is ceded entirely to the participant, and participants are unlikely to be able to find their own thresholds in more than one or two dimensions except if the dimensions have some separable structure that allows them to be adjusted independently. Consequently, classical psychophysics has rarely exceeded one or two stimulus dimensions, and is time-consuming even in that setting.

\subsection{Adaptive parametric methods in psychophysics}

To address the above problems, several adaptive techniques have been developed over the years with the goal of
acheiving similar accuracy with fewer trials \citep{Leek2001}. They make do with less data by injecting
additional structure into the problem, typically a model of the psychometric transfer function, and then they
use some secondary objective or heuristic to determine a sequence of points to sample, either a priori or
conditioned on the data observed so far. The most well-known such methods are PEST, \citep{Taylor1967}, QUEST
\citep{Watson1983}, and Psi \citep{Kontsevich1999}, though others exist as well (\cite{Levitt1971,Watson1983};
see also \cite{Treutwein1995} for a review). It is notable that the most well-used methods are also the ones where robust public implementations are readily available. Nonetheless, most prior adaptive methods either make no explicit assumption about the psychometric
function (in the case of some heuristic methods), or share the assumption of a parametric model for the
psychometric function consistent with Weber's law. They have been applied to various domains in perception including auditory filters \citep{Shen2012}, contrast external noise functions \citep{Lesmes2006}, visual fields \citep{bengtsson1997new}, as well as complex visual models \citep{DiMattina2015}.

These parametric models are still one-dimensional and assume that the stimulus varies on only one dimension.
To evaluate a multidimensional space (for example, the threshold on visual stimulus contrast as a function of
size and color), the additional dimensions are once again evaluated independently over a grid of points. One notable exception to this work is QUEST+ \citep{Watson2017}, which supports multidimensional parametric models.
However, QUEST+ requires the researcher to specify a parametric form for the psychometric function a
priori, such as assuming that the contrast threshold is linear in stimulus size.

By using strongly constrained models during both the experiment stage and the subsequent analysis, these methods strongly constrain the set of conclusions that can be drawn: under model misspecification (i.e. if data violate
some assumptions of the chosen parametric model), the method of constant stimuli would still allow estimation of the correct
function, whereas a restrictive adaptive method would not. The problem is especially acute because these modeling assumptions must be made prior to collecting any data, even in an exploratory setting.

\subsection{Nonparametric models and methods for psychophysics}

Recent work has attempted to address the issue of adaptive methods making strong assumptions about the shape of the psychometric function. Specifically, this work has modeled the psychometric transfer function using a Gaussian process (GP), a Bayesian nonparametric model.

GPs have a long history in sample-efficient modeling of complex functions, and are used to support adaptive sampling in a variety of domains, including geophysics, materials development, genomics, and others (for some reviews, see \citet{Brochu2010ATO,Frazier2018,Deisenroth2015}. They have additionally seen a surge of recent interest and advancement due to their use for global optimization of machine learning hyperparameters \citep[e.g][]{Snoek2012,Balandat2020}. GPs are most commonly used with continuous outcome spaces using a simple Gaussian observation density, but have also been applied for discrete observations using a link function and non-Gaussian observation likelihoods, similarly to the generalized linear model.

GPs have been used to model psychophysical response data in both detection \citep{Song2018,Song2017b,Gardner2015a,Schlittenlacher2018,Schlittenlacher2020} and discrimination \citep{Browder2019} experiments. In the detection work, the GP models were further used for adaptive stimulus selection. In both cases, the response was modeled as a data-driven nonlinear function of multi-dimensional stimuli. In this way, using adaptive sampling with GPs for psychometric experimentation addresses the key issues with both classical methods (sample efficiency) and parametric adaptive methods (strong parametric assumptions).

We build on this prior work in a number of ways. First, we show that the specific assumptions made about the
psychometric function in prior nonparametric work can be improved. Work by Song and colleagues assumed the function is linear, as in conventional psychophysics, and only shifted by context variables (i.e.\ without changing the slope). On the other hand, work by Browder and colleagues let the psychometric function be any smooth function including those with a nonmonotonic relationship between stimulus and perception, which is not a realistic outcome in many psychophysics experiments. We propose a middle ground, a prior over \emph{smooth monotonic functions} that is able to encode known monotonicity without having to otherwise specify the shape of the psychometric function.

Second, we develop new adaptive stimulus selection policies that improve on the prior work by being more tailored to the objectives of psychophysics researchers. We show that the threshold-estimation objective of classical psychophysics can be framed as \emph{level set estimation} (LSE) in multiple dimensions \citep{Gotovos2013}, and provide a new LSE objective to complement the global variance- and entropy-based objectives used in prior work. This gives researchers the ability to tailor their adaptive procedure to their experiment goals, using the LSE objective for threshold estimation or a global objective for estimating the full psychometric field.

Finally, we make explicit the connection between probit-GP models as we use them here and classical psychophysical theory, showing how they can be thought of as a formal generalization of both the Weber-Fechner law and classical Signal Detection Theory.

\section{Theory}

\subsection{Classical psychophysics and signal detection}

The fundamental objective of psychophysics is to quantify the relationship between the properties of stimuli and the perceived sensation, that is, to find a function $f: \mathbb{R}^k\rightarrow \mathbb{R} $ that maps from stimulus value $x \in \mathbb{R}^k$ to the value of a latent percept. In practice, we are not able to measure the outcome of $f$ directly, and instead measure some indirect consequence of it, such as a response from a participant or a neural signal. This is also known as the \emph{transducer function} \citep{May2013}. Thus, to study psychophysics we must estimate not just $f$ but also some additional function $p(y\mid f(x))$ that maps from the latent percept to the probability of some response $y$. This function is typically stochastic, and motivated by some assumption about the distribution of noise somewhere in the transmission chain from stimulus to decision.

The classical empirical finding, established by Weber and formalized by Fechner \citep{fechner1966elements}, is that the minimal physical difference in stimulus intensities that produces a detectable change in sensation (i.e.\ a JND) is approximately a fractional increment. For example, to perceive the difference in two weight stimuli with 0.75 probability, one weight must be 2\% greater than the other. We can express such a relationship as follows:

\begin{align} \label{eqn:fechner}
\frac{df(x)}{dx} &= \frac{1}{xk},
\end{align}

where $k$ is a scaling constant (the Weber contrast). We can now integrate, giving us $f(x) = \frac{\log(x)}{k} + c$ (though see \citet{Dzhafarov2011} for a more careful derivation). This log-linear transformation is fully specified by determining the two constants, $k$ and $c$.
As noted above, we must also specify a response model for the participants. In the standard setting for discrimination between two intensities $x_1$ and $x_2$, the full response model is the following:
\begin{align}
\tilde{f}(x_1) &= \frac{\log(x_1)}{k} + c + \epsilon_1\\
\tilde{f}(x_2) &= \frac{\log(x_2)}{k} + c + \epsilon_2\\
\epsilon_1, \epsilon_2 &\sim \mathcal{N}(0, \sigma^2),
\end{align}
where $\epsilon_1$ and $\epsilon_2$ are independent Gaussian noise terms. $\tilde{f}(\cdot)$ represents the latent percept, and the subjective response is made by comparing them for the two stimuli:
\begin{align}
p( x_1 > x_2 | \tilde{f}(x_1), \tilde{f}(x_2)) = \Phi\left(\frac{\tilde{f}(x_1)-\tilde{f}(x_2)}{\sqrt{2\sigma}}\right),
\end{align}
where $\Phi(\cdot)$ is the standard normal cumulative distribution function (CDF), also known as the \emph{probit}.
From this, it can also be reasoned that other assumptions about observation noise yield other (non-probit)
transformation of the log-intensity difference. Note that this model is not just a formalization of the
classical Weber-Fechner model, but it is also equivalent to the formulation of the problem under classical
signal detection theory \citep{Green1966}, with the term inside the probit equivalent to $d'$. It can also
be interpreted as the likelihood of the popular diffusion decision model with a fixed response time, the
so-called ``interrogation paradigm" \citep{Bogacz2006}, with the term inside the probit being the
normalized drift rate scaled by time (though neither of these other models has an explicit logarithmic relationship inside the probit).

While the Weber-Fechner law has wide empirical support, especially for moderate values of $x$, such a rigid formulation has some disadvantages. In particular, the log-linear relationship breaks down at very high and low stimulus values where the neural response is either not well-defined or fails to saturate. To address this issue, the standard approach is to independently measure absolute detection and discrimination thresholds, i.e.\ implicitly assume a piecewise linear model with $k_{detection}\ne k_{discrimination}$ \citep[e.g][]{Aguilar1954,Legge1984,Foley1981,Mikkelsen2020}.

Furthermore, it presupposes that the stimulus intensity $x$ is available in the perceptually `correct' units,
i.e.\ ones being presented to the observer's sensors, such that the logarithmic transformation can apply. In practice, this has meant the need to manually search over the space of transformations to correct units, estimate a piecewise linear model with multiple slopes and verify that they are all of similar value, or both. While this is less of a concern for low-level visual and auditory stimuli where the intensity dimension is well-understood, it is of greater relevance to more complex intensity dimensions such as gloss \citep{Chadwick2015}, transparency \citep{Beck1984}, or roughness \citep{BergmannTiest2007}, where finding the units over which the log-linear relationship holds is nontrivial.

Next, the globally linear shape of the function means that estimates of the slope (or JND) in the middle of the intensity domain is highly sensitive to outlier values at the edges of the domain, such as those caused by lapses and guesses on the part of the subject. As a consequence, special care is typically taken to model those empirical phenomena \citep{Prins2012,Wichmann2001,Linares2016}.

Finally, this model is one-dimensional, and does not make predictions of multidimensional stimuli except if they can be transformed to a one-dimensional intensity variable that can be transformed to $d'$. This has meant that practitioners interested in multidimensional psychometric fields can either estimate independent one-dimensional slices through the psychometric field, or write explicit parametric models for detection or discrimination thresholds $k(z)$ as a function of additional contextual variables $z$ \citep[e.g.][]{Watson2017}.

\subsection{Standard empirical methods in context of the standard model}
We now revisit standard experimental methods in context of the above formalization. In the classical approach, whether responses are collected by the methods of constant stimuli or limits, the latent perceptual value is never estimated. Instead, each stimulus pair is repeated many times, such that average accuracies can be estimated. Then, a psychometric function in probability space can be fit directly to these probabilities, and the various thresholds estimated from it. In the case of heuristic staircases such as PEST \citep{Taylor1967}, the threshold is read out from the final step of the staircase. In parametric model-based methods such as QUEST \citep{Watson1983}, the threshold can be estimated from the model used. In both cases, the fact that the standard Weber-Fechner model does not simultaneously apply at sub- and supra-threshold levels means that while a psychometric function can be fit to staircase data, it would not provide good estimates of the psychometric slope or JND.

\subsection{Towards nonparametric psychophysics}

We return the Weber-Fechner formulation (Eq.~\ref{eqn:fechner}), to reflect on the desiderata for a more general nonparametric psychophysics: what is the minimal set of assumptions that one should make about the relationship between external stimulus and internal percept? We argue that the primary requirement is for the relationship to be monotonic in the intensity dimension, i.e.\ for increases in external stimulus to drive increases (or zero change) in the internal percept. Formally this gives the following generalized Fechnarian law:
\begin{align}
\frac{df}{dx} &= g(x),
\end{align}
where $g: \mathbb{R} \rightarrow \mathbb{R}^+$, though in practice we operate with the psychometric transfer function
$f(x)$ directly.
For a multidimensional stimulus $x \in \mathbb{R}^k$, this generalized law applies to the partial derivative with respect to the stimulus intensity dimension, and not necessarily other dimensions that describe properties of the stimulus with respect to which perception may be nonmonotonic.

We take as our model of $f(x)$ a Gaussian process (GP) prior. A GP is a stochastic process that defines a distribution over functions (in our case, psychometric fields) $p(f(x))$, such that the distribution of any finite set of values measured from these functions (in our case, the psychometric field evaluated at a set of points) is jointly multivariate normal.

A GP is described by two functions, a prior mean function $\mu_{\theta}(x)$ and a kernel function $K_{\theta}(x, x')$ that defines the covariance in function values for any pair of points in the parameter space: $\textrm{Cov}[f(x_1), f(x_2)] = K_{\theta}(x_1, x_2)$. The kernel is typically constructed so that the covariance between function values is high when $x_1$ and $x_2$ are close, and low when $x_1$ and $x_2$ are far apart, which induces smoothness in the function across the parameter space: similar stimuli will produce similar outcomes, whereas disparate stimuli will not affect each other. With a GP, the posterior for the function values at any collection of points will be a multivariate normal whose mean and covariance can be written in closed form (and will in general depend on the covariance function and prior mean). As with other kernel-based methods such as kernel density estimation, appropriate choices of these two functions can yield priors that prefer simpler models when data are limited while still converging to any true function in the limit of data. It is also straightforward for the mean and kernel functions to operate over multidimensional inputs (this is, in fact, the standard setting for GPs), making this generalization both multidimensional and potentially nonlinear. See \citet{rasmussen2006gaussian} for a full review of GPs.

The observation model remains the same as in the classical setting, with the assumption of Gaussian noise giving way to a probit transformation. Alternate transformations such as the logit or Weibull CDF are also possible, but unnecessary if the underlying underlying psychometric model is flexible enough, as a GP is. GPs perform a local interpolation such that estimates in the middle of the intensity domain are less sensitive to values at the edges than a global linear model, making them more robust to guesses and lapses. This formulation thus addresses the theoretical concerns raised above: an $f(x)$ modeled by a GP is flexible enough to model correct saturating behavior at high and low intensities, can model incorrect scaling of stimulus intensity, and is multi-dimensional by default.

A closely related approach was given by \citet{Song2018} and \citet{Browder2019}, who also model the psychometric function using a GP. Browder and colleagues consider the general case where both $x_1$ and $x_2$ are allowed to vary, whereas Song and colleagues restrict one of the stimuli (e.g.\ $x_2$, without loss of generality) to be a constant psychophysical standard. We follow the latter, which makes the latent intensity of the standard $f(x_2)$ a constant and reduces the model to a GP classification model over the value of the remaining stimulus. To respect the monotonicity assumption on $f$ in the intensity dimension, Song and colleagues use a linear kernel in that dimension, with an added radial basis function (RBF) kernel in the context dimensions. This provides a multidimensional model with flexibility on contextual effects but does not address scaling and nonlinearity issues in the intensity dimension, and assumes that the effect of contextual dimensions is additive.

In contrast, we use an RBF kernel on all dimensions, using a constraint on the derivatives of the GP to maintain monotonicity in the intensity dimension \citep{Riihimaki2010}. This allows us to consider more flexible psychometric functions both in the intensity dimension and in how the intensity dimension interacts with remaining dimensions. We additionally introduce a new acquisition function for adaptive psychophysics, as discussed below.

\subsection{Extracting slopes and thresholds from multidimensional nonparametric models}
After estimation of the latent psychometric function $f(x)$, there are two possible definitions of a JND or
discrimination threshold. The first takes the formal definition above literally, and defines it as the derivative of the psychometric field at a point $g(x) := f'(x)$. A second definition takes a discretized view, which is that the JND is the change in $x$ needed to increase $f(x)$ by one standard deviation, that is, the smallest stimulus increment $D(x)$ such that $x$ and $x+ D(x)$ can be distinguished with probability at least $\Phi^{-1}(x)$. Alternatively, this can be formulated as $f(x+ D(x))  \ge f(x)+ 1$. The definitions coincide for a linear model of $f(\cdot)$, and JND under either definition can be extracted from the model-estimated $f(\cdot)$ in the nonparametric setting. This can be done by finite-differencing in the case of the derivative definition, or locally-linear interpolation for the step-JND.

\subsection{Gaussian processes with monotonicity information}
\begin{figure}[htb]
    \centering
    \includegraphics[width=\columnwidth]{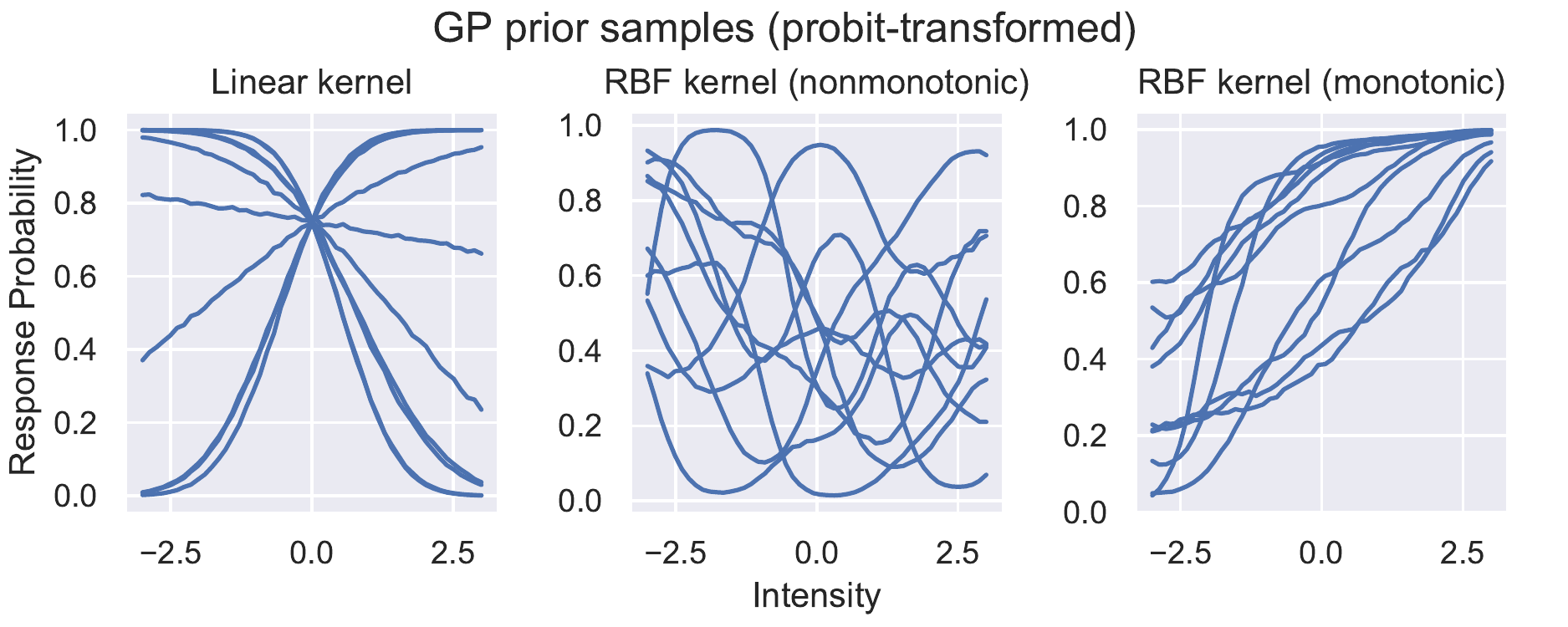}
    \caption{\textbf{Prior samples from the three GP models (in 1d).} \emph{Left:} the linear-additive model's prior is fairly restrictive, but also includes negative slopes (though these are quickly discarded from the posterior after a handful of observations). The additive context (not shown) in $f$-space turns into a shift in probability space, but the posterior slope is restricted to be the same over all contexts. \emph{Middle:} the conventional RBF model is very flexible, including functions that are far from monotonic. \emph{Right:} the approximately-monotonic RBF model is more flexible than the linear kernel model, but excludes the nonmonotonic functions from the prior seen in the other two models. }
    \label{fig:prior_samps_fig}
\end{figure}

To restrict our model to be monotonic in the intensity dimension, we follow \citet{Riihimaki2010} in exploiting the fact that the derivative of a GP is itself a GP. This means that the joint distribution of any set of observations from $f(x)$ and $f'(x)$ is a multivariate normal distribution with closed form mean and covariance. We produce a monotonic prior for $f(x)$ by conditioning on a restriction that $f'(\tilde{x}) \geq 0$, at a set of derivative inducing points $\tilde{x}$. In doing so, we make no assumption about the shape of $f(x)$ other than that its derivative is non-negative, and we are thus able to directly encode the generalized Fechnarian law above. Fig. \ref{fig:prior_samps_fig} gives an illustration of prior samples from this model, along with those from prior work that were described above (the linear-additive kernel, and a full RBF kernel). The prior samples show clearly the restrictiveness of the linear kernel and the unrealistic non-monotonic functions produced by a full RBF kernel. Our monotonic GP is able to encode known monotonicity without sacrificing important flexibility in understanding the relationship between the stimulus and context variables.

\subsection{Inference and acquisition}

\begin{figure}[htb]
    \centering
    \includegraphics[width=\columnwidth]{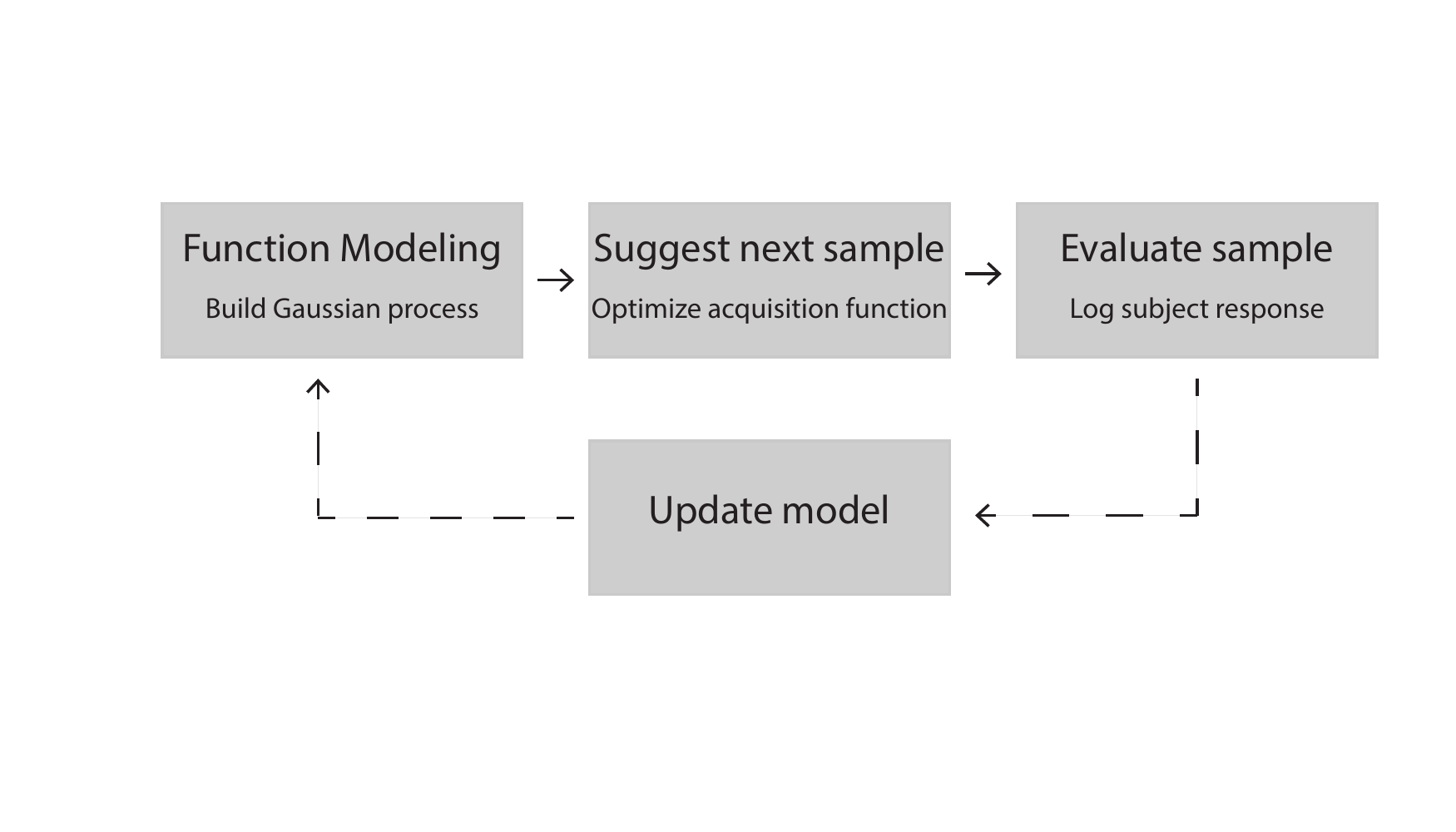}
    \caption{\textbf{The active sampling loop.} In our method, the experiment proceeds by first modeling the data using a Gaussian process, optimizing an acquisition function to determine the next stimulus to sample, presenting the stimulus to the participant and collecting the response, and then incorporating that new data into the model to continue the process.}
    \label{fig:flow_chart_fig}
\end{figure}

A significant advantage of using models for estimating the psychometric function is that it enables active learning and adaptive sampling, which can greatly improve sample efficiency and reduce the time needed to run an experiment. Fig.~\ref{fig:flow_chart_fig} provides an illustration of the active learning process as it can be applied to a psychophysics experiment. We first estimate a model of the data so far (the \emph{inference} step), then use that model to optimize an acquisition function that is maximized at the stimulus parameters we should evaluate next (the \emph{acquisition} step), present the stimulus to the participant and then log the data so we can update the model, and the cycle continues. We now describe these steps in detail.

\subsubsection{Inference}

Our need to update the model frequently, with the human in the loop, means that both model inference and selection of the next point must take a few seconds at most. This precludes the use of full MCMC inference, and requires the use of techniques for scalable approximate GP inference. The standard approach to scalable inference for GPs with non-Gaussian observations such as ours is variational inference \citep[e.g.][]{rasmussen2006gaussian}. Variational inference approximates the posterior distribution $p(f\mid y)$ with a simpler distribution $q(f\mid y)$ (often a multivariate normal), chosen to minimize the Kullback-Leibler (KL) divergence between the true and approximate posteriors. This is made tractable by minimizing an objective known as the evidence lower bound (ELBO), and is much faster than performing full MCMC inference (for more on variational inference, see \cite{Blei2017}). However, variational inference is difficult to implement with monotonicity information because the approximate posterior $q(f\mid y)$ needs to be monotonic, which can render the ELBO intractable.
Consequently, previous approaches to monotonic GPs used expectation propagation \citep{Riihimaki2010}.

Rather than implement custom inference methods for the various models we consider, to maximize simplicity and scalability we use conventional variational inference throughout. In the monotonic case, we do this by performing conventional, non-monotonic inference, and then performing rejection sampling on the non-monotonic posterior to construct an approximately monotonic posterior: we draw a large set of samples from the GP, and attempt to exclude all samples with negative derivative values.
To avoid long delays we draw a fixed number of samples, in which case rejection sampling does not guarantee that we will achieve the desired number of monotonic posterior samples. In that case, we draw the samples that least violate the monotonicity constraint by including those samples with negative derivatives that are closest to 0. This produces a posterior that is nearly monotonic, and as the variational posterior converges with increasing data, the rejection rate in the sampling goes to zero assuming the data-generating function is indeed monotonic.
\subsubsection{Acquisition}
\begin{figure*}[htb]
    \centering
    \includegraphics[width=\textwidth]{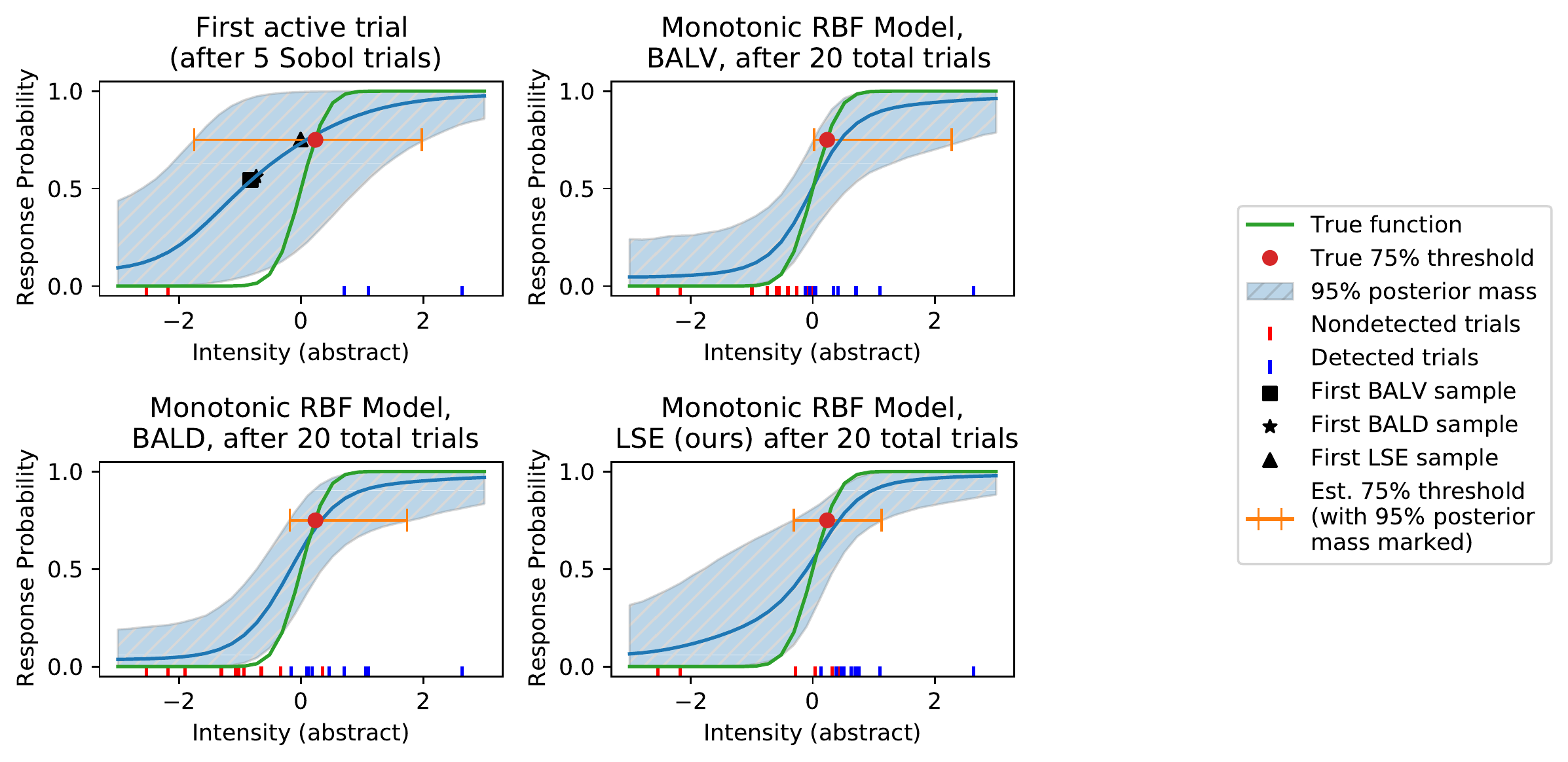}
    \caption{\textbf{Different acquisition functions can target different experimenter objectives.} \emph{Top left}: The posterior over a simple linear psychometric function after 5 Sobol trials, and the first adaptive trial selected under BALD, BALV, and LSE targeting the 75\% threshold. LSE samples where the current estimate would place the threshold, whereas BALD and BALV both sample close to 0.5, where uncertainty is maximized and the most information about the function can be gained. \emph{Top right, bottom left, and bottom right}: The same estimates after 15 additional adaptive trials, showing that while BALV and BALD sample across different values of the function and therefore provide a better estimate of it overall, LSE samples near the eventual threshold, achieveing a tighter final interval. The monotonic RBF model is used for all three acquisition functions.}
    \label{fig:same-model-different-acq}
\end{figure*}

Our primary purpose in inserting a model into the experimentation process is to use active learning to reduce the number of observations required to determine the psychophysical quantities of interest. The two ingredients of such an active learning process are the model (the GP just described), and a policy that determines how to choose the next stimulus to present based on the current model estimates.
Conventionally in GP active learning, such a policy is encoded in an \emph{acquisition function} $a(x)$ that maps from the stimulus parameters to the desirability of sampling in that location. By choosing different acquisition functions, it is possible to make the active learning process reflect different goals, for example estimating the JND vs.\ estimating the detection threshold (see Fig.~\ref{fig:same-model-different-acq} for an illustration of this point). The next point can be selected by choosing the point where the acquisition function is maximized, or selected randomly in proportion to the value of the acquisition function (the latter being an instance of Thompson sampling; \cite{Thompson1933}). Using Thompson sampling instead of solving the minimization problem on $a(x)$ can provide speed benefits, as well as encourage model exploration. A final baseline strategy is to use a random or quasi-random sequence over the search domain. In practice, we begin the experiment with a fixed number of trials drawn from a quasi-random sequence (in our case, a Sobol sequence; \cite{Sobol1967}), after which we sample points according to our acquisition function. We consider three acquisition functions below: mutual-information maximization (also known as Bayesian Active Learning by Disagreement or BALD; \cite{Houlsby2011}), posterior variance minimization (also known as Bayesian Active Learning by Variance or BALV; \cite{Settles2009}), and level set estimation (LSE; \cite{Gotovos2013}). Note that while BALD and BALV have been used in psychometrics previously, the introduction of LSE is new, and is motivated by the problem of threshold estimation. We discuss all three acquisition functions next.

\paragraph{Bayesian active learning by variance (BALV)}
If the objective is to estimate the full psychometric function, one natural solution is to simply attempt to minimize the uncertainty over the whole psychometric surface, by sampling wherever the current posterior variance is highest. This strategy has been previously used for estimation of psychometric fields and thresholds \citep{Song2018,Song2017b,Gardner2015a,Schlittenlacher2018,Schlittenlacher2020}. The probit transformation means that even if variance is very high in some parts of the space, that variance cannot be reduced if the response probability there is squashed to 0 or 1. Thus, BALV is conventionally performed with respect to the variance of the response distribution, i.e.\ $Bernoulli(\Phi(f(x))):=\Phi(f(x))(1-\Phi(f(x)))$. If uncertainty over $f(x)$ is uniform over $x$, BALV reduces to sampling wherever the probability of response is closest to 0.5 (though in practice sampling is slightly more variable due to variations in the variance of $f(x)$). This is somewhat undesirable if one is interested not in the 0.5-threshold, but in another threshold such as 0.75. It is also greedy: repeatedly sampling at the point of highest current uncertainty is not equivalent to sampling at points which will maximally reduce total, global uncertainty.

\paragraph{Bayesian active learning by disagreement (BALD)}
Related to the issue brought up above, minimizing the uncertainty over the full surface may be better accomplished by sampling a point which will most reduce the posterior entropy, in expectation over outcomes. This acquisition function is known as BALD in the GP literature \citep[e.g.][]{Houlsby2011}, and is equivalent to the mutual information between the unseen outcome $y$  and the response probability $\Phi(f(x))$. In the psychophysics literature it is among the most common acquisition functions, used in both nonparametric models (\emph{Ibid.}) and more classical adaptive methods \citep[e.g.][]{Watson2017,Kontsevich1999,Brand2002,DiMattina2015,Hall1981,Watson1983,Kujala2006,Lesmes2010}. BALD is less greedy than BALV (using effectively one-step lookahead) and is more likely to sample over the full domain of the function. While this may be desirable for estimating the full psychometric surface or the JND, this is less desirable when the experiment objective is to target a specific part of the surface.

\paragraph{Level Set Estimation (LSE)}
When the goal of an experiment is threshold estimation, we would ideally want to focus sampling at the places that most improve our estimate of the threshold. When the function is monotonic with respect to stimulus intensity, the detection threshold is a point in one dimension, forms a line in 2d, a plane in 3d, and so on. Formally, it is called a \emph{level set} of $f(x)$, and is the set of all points $x$ such that $f(x)=T$ for some desired threshold level $T$. Our goal in psychophysical threshold estimation is to identify the threshold stimulus intensity for any given value of the contextual variables, which corresponds to identifying the entire level set across the parameter space.

\citet{Gotovos2013} proposed an acquisition function for level-set estimation, called the LSE acquisition function, that selects points according to their ambiguity of being either above or below threshold. This provides a natural approach for active sampling to reduce uncertainty in the threshold intensity. They show how the ambiguity measure can easily be computed directly from the posterior mean and variance of $f(x)$, though in contrast to this prior work we use $\Phi(f(x))$ instead, which gives markedly better performance in our setting. To select the next stimulus to display, we use standard gradient optimization techniques to optimize the LSE acquisition function and identify the maximum ambiguity point.

We also develop and explore a novel Thompson-sampling variant of LSE (LSETS) which, instead of optimizing the acquisition function, takes a large, joint sample from the GP posterior on a space-filling design and selects the next stimulus $x$ according to its probability of its $f(x)$ being closest to $T$. This improves running time by avoiding solving an optimization problem for each sample selection.

\section{Methods}

\subsection{Models}

Our primary model for the experiments in this paper is the monotonic probit RBF-GP model described above. We also evaluate the same model without the monotonicity constraint in order to independently evaluate the benefit of monotonicity relative to the benefit of our novel acquisition function. To provide a fair comparison, this nonmonotonic model uses the exact same variational inference and rejection sampling procedure described above, except we accept all samples. Finally, we include an implementation of the separable linear-RBF model of Song and colleagues also described above, as this is the most closely related prior work to ours. Since no public implementation of this model is available to our knowledge, we use our own implementation, with one difference: we use variational inference (VI) rather than expectation propagation (EP). The use of a consistent inference algorithm makes for a fairer comparison with our models, but VI and EP can perform differently in different settings. Likely due to this change in inference approach, our results do not perfectly replicate that reported in the prior work.

GP kernels typically incorporate several hyperparameters that are fit to the data, typically to maximize marginal likelihood of the data. For the RBF kernel, the hyperparameters are the \emph{length scale}, which determines the distance in stimulus space for which points remain correlated, and the \emph{output scale}, which is an overall scale on function values. The linear kernel used in the additive model also has an output scale. As is standard practice in GP regression, we gave priors to these parameters to regularize the fitting and improve inference. Our objective for the output scale was to encourage the values of the latent psychometric field $f(x)$ to be in the operating range of the probit function, so we used a uniform prior between 1 and 4 (corresponding to covering response probabilities spanning a range of 0.84 in the lower end or 0.99997 in the upper end). Our objective for the lengthscale was to exclude lengthscales larger than the range of the data (since they are unidentifiable) or smaller than 10\% of the range of the data (since such highly irregular functions are unlikely in psychophysics). We thus used an inverse-gamma prior with parameters set such that 98\% of the probability mass was within this range.

\subsection{Optimizing the objective function}

For the nonmonotonic and monotonic full-RBF GPs, we optimize the acquisition function by stochastic gradient descent, since the use of rejection sampling makes the posterior stochastic and precludes the use second-order gradient methods that assume noiseless function evaluations. For the linear-additive model we include an additional heuristic proposed by \citet{Song2018} of adding noise to encourage exploration. For this heuristic, we first compute the acquisition function over a dense grid of points generated from a quasi-random Sobol sequence, normalize the values to be between 0 and 1, add a $\mathcal{N}(0, 0.2)$ random value to each, and pick the max of the noisy value. We confirmed that introducing this heuristic indeed helped to somewhat mitigate the boundary over-exploration effects we otherwise saw in that model (though as we will see below, these issues were not fully mitigated).

\subsection{Test functions based on audiometry data}
\begin{figure}[htb]
  \centering
  \includegraphics[width=0.95\columnwidth]{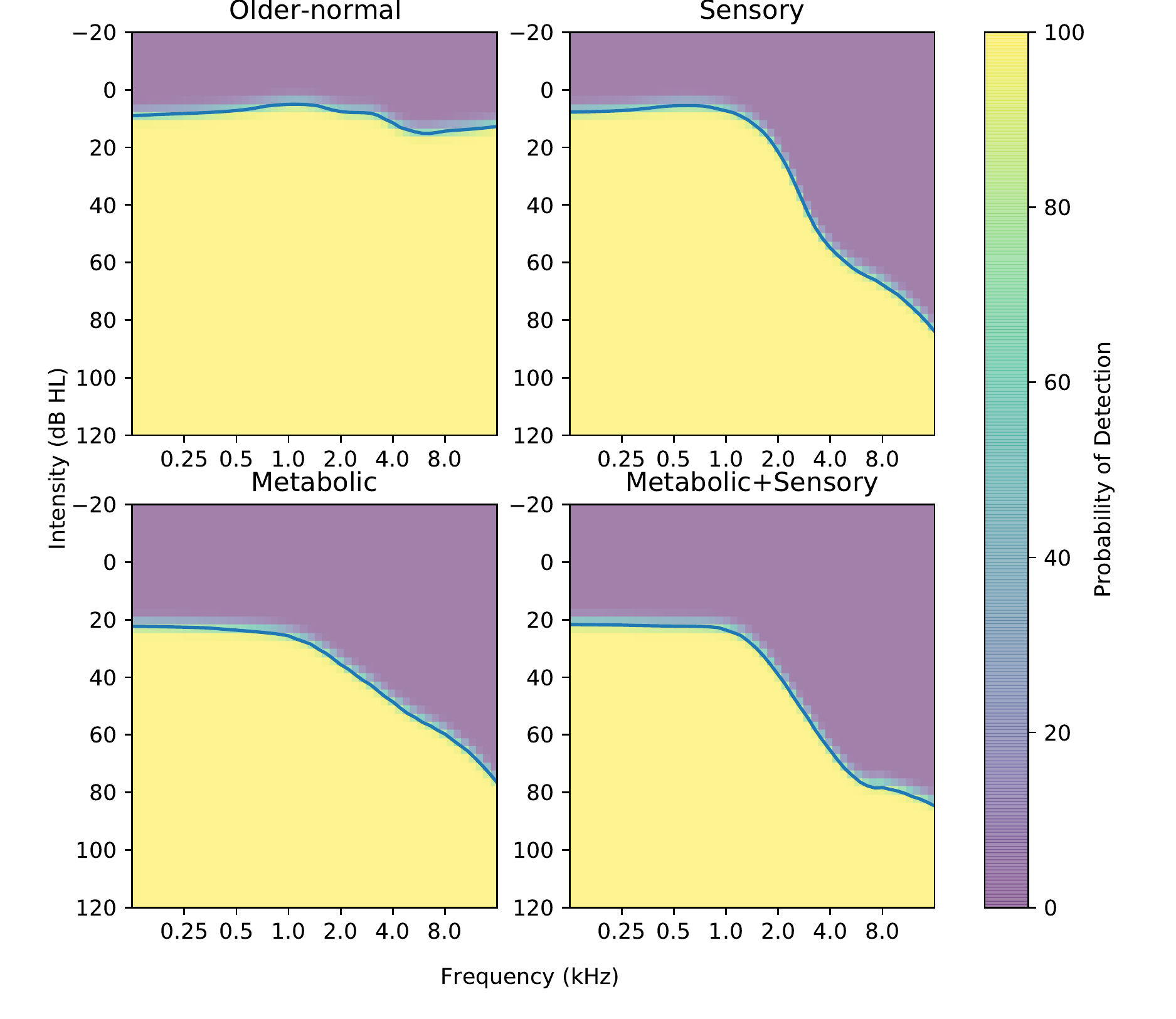}
  \caption{\textbf{Audiometric test functions ground truth}. Surfaces generated with \protect{$\beta=2$} for the $f_{song}$ function described in the main text.}
  \label{fig:audiometric-thresholds}
\end{figure}

Previous applications of nonparametric psychophysics have focused primarily on the domain of psychoacoustics, specifically pure tone auditometry. To compare to this past work, we used the same set of test functions: the mean thresholds of 338 exemplars of four audiometric phenotypes defined from an animal model of age-related hearing loss. In all cases they define a hearing threshold in dB as a function of tone frequency (in kHz), measured from pure tone audiograms measured using standard audiometric procedures. Since no public implementations of these test functions is available, we followed the same procedure as \citet{Song2017b} to generate our own version: we first extracted ground-truth audiometric thresholds from Fig.~2 in \citet{Dubno2013}, and used cubic spline interpolation and linear extrapolation to evaluate them over the full range from 0.125 to 16 kHz\footnote{We thank the authors for their clarification by email about this procedure}. Then, to evaluate the latent psychometric field at any point we first interpolate/extrapolate the threshold from those measured values, and then use the standard linear model with that threshold. That is, $f_{song}(x_c, x_i) := \frac{x_i-\theta_{song}(x_c)}{\beta}$, where $x_i$ is the intensity value (in dB), $x_c$ is the context value (in kHz), $\theta(\cdot)$ is the interpolated threshold and $\beta$ is a psychometric spread parameter. As in previous work, we use all four phenotypes from \citet{Dubno2013} as test functions, and a grid of $\beta \in [0.2, 0.5, 1, 2, 5, 10]$. We plot the ground-truth threshold functions in Fig.~\ref{fig:audiometric-thresholds}.

\subsection{New test functions for adaptive psychophysics}
The above test function, while based on real psychometric thresholds, makes a strong separability assumption between context and intensity: it assumes that the context does not interact with intensity but simply shifts the psychometric function. Furthermore, the empirically informed thresholds have fairly simple relationships with the contextual variable, all taking a sigmoid shape with large flat portions where context has minimal effect on the threshold. We introduce a novel, more difficult parametric test function for adaptive psychophysics, in two variants. First, we replace the interpolated context function with a fourth-order polynomial:

\begin{align}
  \theta_h(x_c) := 2(0.05 + 0.4(-1 + 0.2x_c)^2 x_c^2),
\end{align}

\begin{figure}[htb]
  \centering
  \includegraphics[width=0.95\columnwidth]{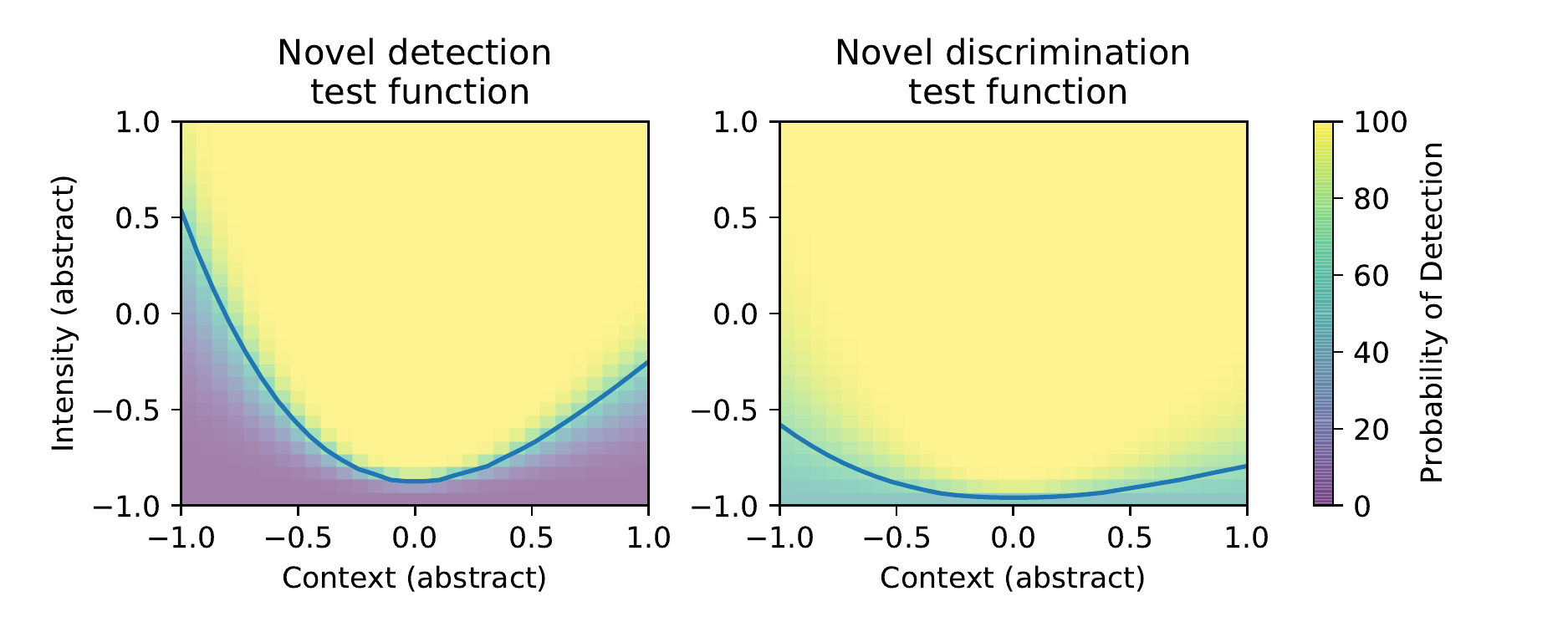}
  \caption{\textbf{Novel test functions for adaptive psychophysics}. On the left is the detection variant (response probability is approximately 0 at the lowest intensity), and on the right is the discrimination variant (response probability is 0.5 at the lowest intensity). The line marks the 0.75 threshold.}
  \label{fig:novel-testfuns}
\end{figure}

Next, we create a simple multiplicative interaction between the context and intensity dimensions. For a detection variant, we use the following transformation:
\begin{align}
  f_{det}(x_i, x_c) := 4\frac{1 + x_i}{\theta_h(x_c)} - 4,
\end{align}
which (after probit transformation) takes on the value of approximately 0 for 0 intensity. This matches standard detection experiments, where participants respond based on the presence of a stimulus. For a discrimination variant, we use:

\begin{align}
  f_{disc}(x_i, x_c) := 2\frac{1 + x_i}{\theta_h(x_c)},
\end{align}
which has response probability of 0.5 for 0 intensity. This matches standard discrimination experiments, where participants respond based on detecting a difference between pairs of stimuli. Both functions are visualized in Fig.~\ref{fig:novel-testfuns} over their input domain of $[-1, -1]$ to $[1, 1]$. In contrast to the audiometry test function, these psychometric functions depend on the context dimension over the entire domain. In addition, while they retain linearity in $x_i$, the interaction with context is now multiplicative, violating the separability assumption. While this test function is not directly drawn from real data, we think it provides a more realistic reflection of the complexities of psychometric data, especially in domains like haptics and multisensory perception.

\subsection{Simulations}
Our simulation set was a full permutation over the following items:
\begin{itemize}
  \item Three model variants: the linear-additive GP model used in past work, a conventional probit-RBF GP, and the monotonic RBF GP.
  \item 26 test functions, consisting of [a] the two variants of our novel test function defined above (detection and discrimination), and [b] 24 audiometric test functions derived from the four phenotypes from Dubno and colleagues and the six noise levels previously used by Song and colleagues (0.2, 0.5, 1, 2, 5, and 10).
  \item Four acquisition functions: BALD, BALV, LSE, and the Thompson sampling variant of LSE (LSETS). We did not evaluate LSETS for the additive GP model because the noisy-acquisition heuristic is conceptually similar. We also included a Sobol acquisition baseline.
\end{itemize}

We initialized each simulation with 5 Sobol trials and ran it for 145 adaptive trials, for a total of 150 trials. In all cases our objective was to estimate the 0.75 threshold. Preliminary experiments with 20 Sobol trials showed the same patterns we report here, though they generally yielded slightly worse performance for the full RBF-based methods and better performance for the additive method (but again, the ordering of results was not changed by this for any test function). We repeated each simulation 100 times under different simulation seeds, for a total of 33800 simulations. The full set of simulations took about a week to run on a single 96-core workstation running on Amazon's EC2 cloud.

We made a conscious decision not to include classical methods: we did not include the method of constant stimuli because even the maximal trial counts we are considering here (150) are insufficient for this method, with trial counts in the 1000s needed for 2d spaces. We did not include parametric adaptive methods because, as noted above, they either have no way of scaling to higher dimensions except by using a grid (QUEST, PEST, etc) or require a parametric assumption for the effect of context that is not available a priori for any of our test functions (QUEST+).

\subsection{Evaluation}
For each simulation, we computed mean absolute error against the ground-truth response probability evaluated over a $30\times 30$ grid. We also evaluated the mean absolute error against the ground-truth threshold. Thresholds were computed by finding the two nearest points to the desired threshold for each value of the context dimension in the aforementioned grid, and performing local linear interpolation between those points to find the threshold. This was done identically to the ground truth function and the model-based estimate. We recorded additional performance metrics (correlation between true and estimated value, maximum absolute error, mean squared error) that showed the same patterns we report here.

\section{The AEPsych toolbox}

We recognize that a number of methods and ideas we introduce in our work are well-known in the statistics and machine learning communities but relatively unfamiliar to psychology researchers (though see \cite{Schulz2018} for one attempt to improve this). This creates a practical barrier to usage of these new adaptive methods in the field. Furthermore, the most popular adaptive methods share the distinction of there being a robust public implementation of the method available. Therefore, an important component of our contribution is \texttt{AEPsych}, a flexible toolbox that supports the usage and further development of these methods for psychophysics.

\subsection{A state of the art base}

The modeling functionality in \texttt{AEPsych} is wrapping and extending \texttt{gpytorch} \citep{Gardner2018}, one of the main state of the art GP modeling packages. The active learning and acquisition components wrap \texttt{botorch} \citep{Balandat2020}, a state of the art package for Bayesian optimization and active learning. Those packages themselves are built on top of PyTorch, one of the dominant software toolkits for machine learning (ML). Building on these toolkits helps ensure not only that we are relying on the current state of the art in GP modeling and Bayesian optimization, but that further advances and extensions in the broader GP and ML communities are automatically inherited by \texttt{AEPsych}. At the same time, we provide out-of-the-box implementations of the models described in this paper as well as the prior linear-additive GP model so that practitioners do not need to interact with these other packages if this is not their research interest.

\subsection{Meeting researchers where they are}

The dependency on the PyTorch ecosystem means that the core modeling components of \texttt{AEPsych} are implemented in Python. However, we recognize that not all researchers work in Python and not all perceptual experiments can be easily implemented in the language. To support the actual way researchers work, \texttt{AEPsych} is able to operate in a mode that allows the interaction with the participant to happen in another language according to the user's preference. To do this, it uses a Python sever for modeling and selection of the stimulus to display next; the server communicates with a client (in any language) that actually displays the stimulus and collects the response. The client and the server can run on the same machine: we use the network interface as a lightweight way to integrate applications in different languages. We provide client implementations for Python, Unity (C\#) and MATLAB to capture the typical ways perceptual researchers work. The client-server interface is very lightweight, consisting of a small number of message types encoded as text in JSON format. We also provide Docker images of the server components, so that users who otherwise do not use Python can have a single monolithic install of our tooling without needing to build a full Python development environment. In addition, we have a lightweight experiment configuration framework using text-based INI configuration files, so an experiment can be configured without editing any Python code, and we provide example configurations for standard experiments. Finally, \texttt{AEPsych} includes a set of tutorials for typical use cases and a large suite of unit tests to ensure correct functionality.

\subsection{Serving the full experimentation use case: pre- and post-experiment}
An important requirement before any experimental data are collected is to understand the amount of data needed. \texttt{AEPsych} provides benchmarking and power analysis tools that practitioners can use to plan their experiments: given a set of assumptions about the shape of the psychometric field and response noise levels, the benchmark module can simulate a full experiment. By repeating this simulation, the researcher can assess typical estimation error and data needs under different assumptions, and use this to design their real experiment. The configuration language for benchmarks and real experiments is identical, so the best benchmark configuration can then be used in a real experiment. The benchmarking module is available in Python only and supports parallelized simulation for efficiency. It was used to generate all of the simulation results in this paper.

An important component after the experiment is data analysis and visualization. \texttt{AEPsych} includes pre-implemented models for analyzing psychophysical data. These models extend what is possible for real-time experimentation by supporting more accurate inference using Markov chain Monte Carlo (MCMC) techniques rather than variational inference, and also provide hierarchical models that can aggregate across participants to generate average psychometric fields while integrating over subject-specific biases. These models are implemented in Stan \citep{Carpenter2017}, a declarative modeling language in which models are close in notation to the underlying mathematical formulation, and which compiles into highly efficient MCMC samplers.

\subsection{A brief overview of the design of AEPsych API}
For flexible implementation of new models, \texttt{AEPsych} has a simple hierarchy of class interfaces. At the lowest level is a \texttt{Modelbridge}, which combines a \texttt{gpytorch} model and acquisition function in one object. Next, \texttt{Strategy} objects describe ways of sampling new observations, which can be based on data the model has seen so far (a \texttt{ModelWrapperStrategy}) or not based on a model (e.g.\ \texttt{SobolStrategy}). \texttt{Strategy} objects can also be composed together, for example creating a sequential strategy that begins with random or Sobol trials and then switches to a full GP model, or a sequential strategy that begins with a simpler model and switches to a more complex model as more data is acquired. Finally, a server object operates a strategy (or set of strategies, for independent interleaved experiments) and logs all data to a local database. The database supports full replay of experiment sessions, and we additionally provide a utility to output the collected stimuli and responses into a text-based CSV file.

\section{Results}

\subsection{Results for audiometric test functions}

\begin{figure*}[!htb]
    \centering
    \includegraphics[width=\textwidth]{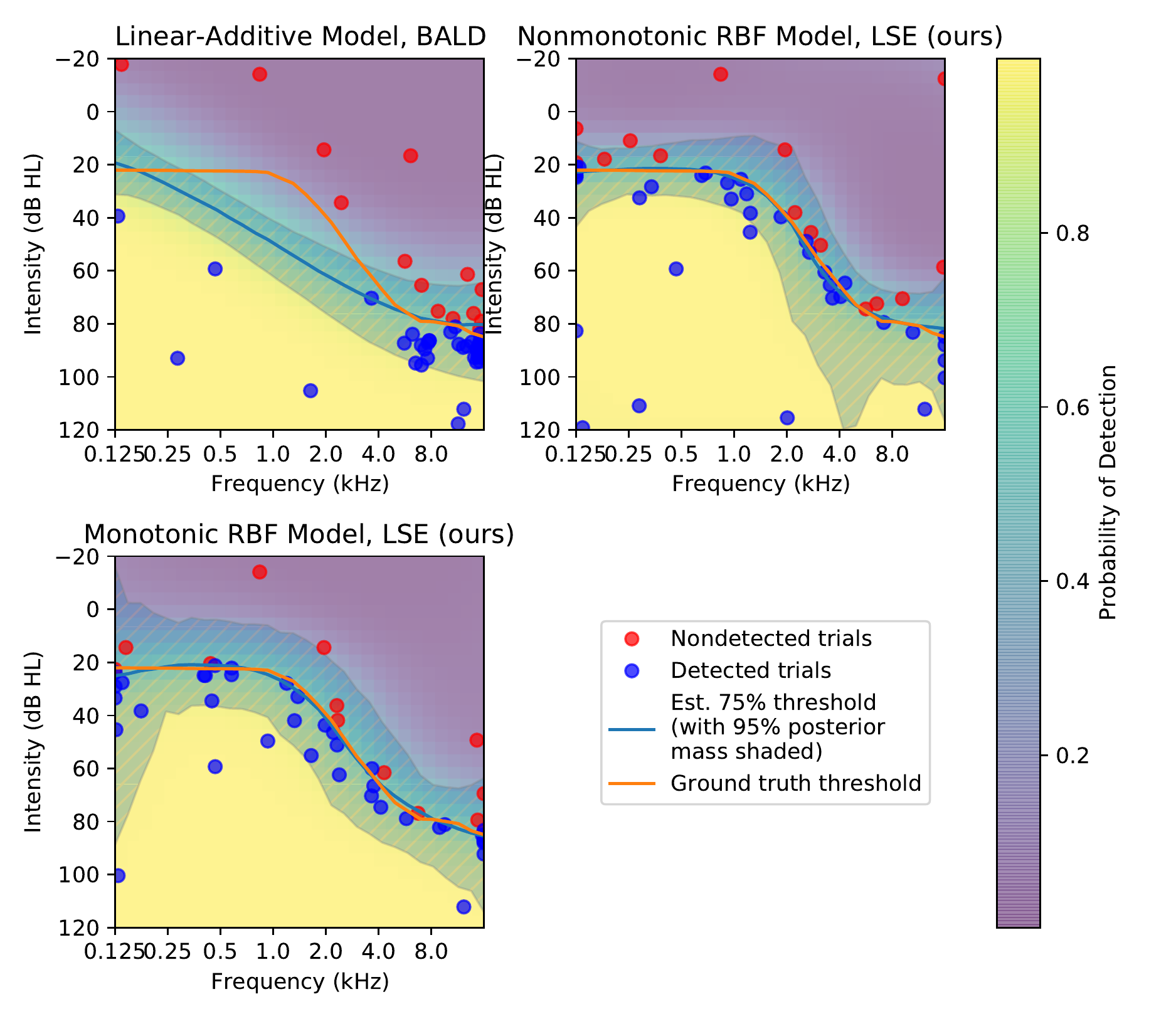}
    \caption{\textbf{Example of samples taken by three representative models on the Metabolic+Sensory test function after 50 trials with $\beta$=2}. All models begin with the same 5 trials generated from a Sobol sequence and then proceed according to their acquisition function (BALD in the case of the linear-additive model; LSE otherwise). Substantial boundary over-exploration is apparent in the linear-additive model, even with the addition of the previously-reported exploration heuristic. Both RBF models with the LSE objective sample consistently at putative threshold locations and produce good mean threshold estimates after only 50 trials.}
    \label{fig:songexample}
\end{figure*}

The four test functions and six psychometric spread values we use add up to 24 total test function variants for the audiometric test function. To understand the broader performance patterns, we first illustrate a sample optimization run from the Metabolic+Sensory phenotype with $\beta=2$ in Fig.~\ref{fig:songexample}, comparing our contribution to prior methods. Under the linear-additive kernel, the model is overconfident in its ability to interpolate across the context dimension, and as a result oversmoothes the threshold and oversamples in one location. In contrast, both RBF models, while still taking some samples at the edges, spend more time exploring the interior of the threshold. Interestingly, the monotonic RBF model samples more at the edges in this specific example than the non-monotonic RBF model, again consistent with a pattern of greater boundary over-exploration being driven by model overconfidence. Finally, we see the RBF model perform similarly with and without the monotonicity constraint, something that we see in aggregate as well.

\begin{figure*}[!htb]
    \centering
    \includegraphics[width=1.1\textwidth]{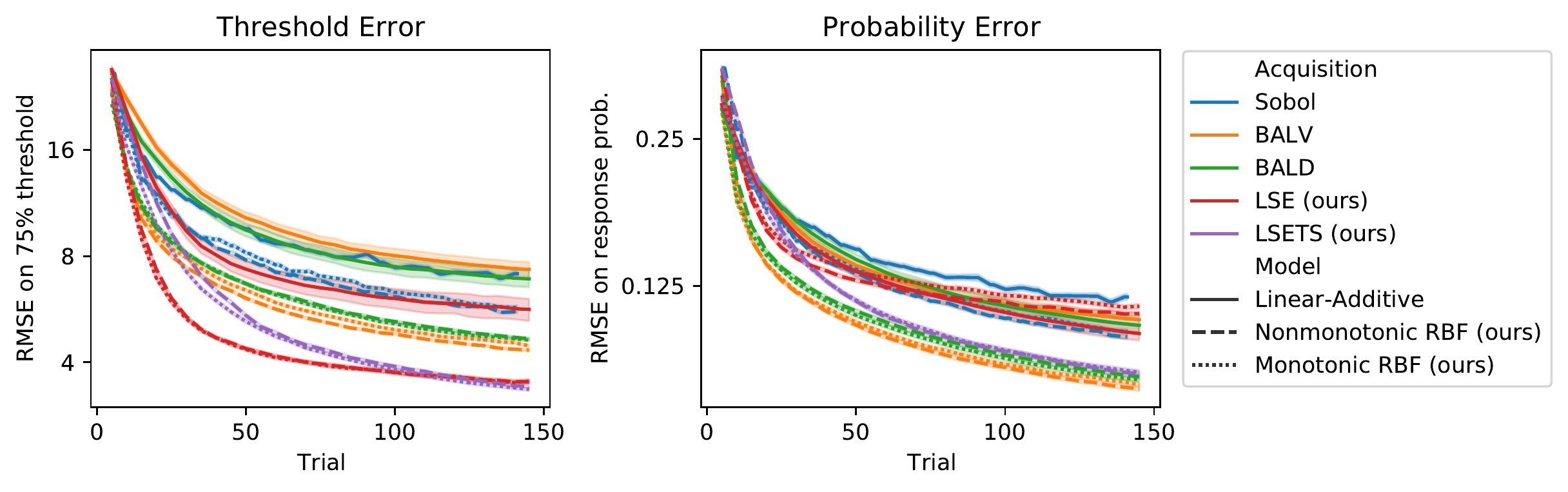}
    \caption{\textbf{Audiometric test function performance}. \emph{Left}: For threshold estimation, LSE consistently the best objective, consistent with its explicit targeting of putative threshold locations, independent of model. \emph{Right}: For estimating the full posterior, BALV appears superior in this setting, followed by all the other objectives excluding LSE, whose focus on the threshold location leads to undersampling of the rest of the surface. In both evaluations, the monotonic and nonmonotonic RBF models are far superior to the linear-additive model, whose average performance is poor overall. Finally, the benefit of monotonicity seems minor at best over the nonmonotonic model. Shaded intervals are a 0.95 confidence interval over simulations. Note the log-scale on the y-axis.}
    \label{fig:songsims}
\end{figure*}

Next, we turn to the aggregate performance curves in Fig.~\ref{fig:songsims}. In the aggregate data, we see that the RBF kernel models we introduced consistently outperform the previously-reported linear-additive kernel models, for both threshold estimation and estimation of the full surface. In addition, we see the LSE acquisition function we introduced performed best for estimating the threshold, in the sense that it both reduced error the fastest and achieved the lowest error (occasionally tied with BALV/BALD) after 150 trials. This confirms the benefits of an acquisition function explicitly designed to target where the threshold is likely to be rather than sampling to reduce global uncertainty: in some cases LSE achieves the same error in well under 50 trials that BALD achieves in 150 trials. This comes with a tradeoff against estimating the full surface which the LSETS acquisition objective we introduced mitigates somewhat. Importantly, the improvements provided by our contributions can be seen even in this setting, where test functions are generated in accordance with the additive structure in the linear-additive model.

\begin{figure*}[!htb]
    \centering
    \includegraphics[height=\textheight]{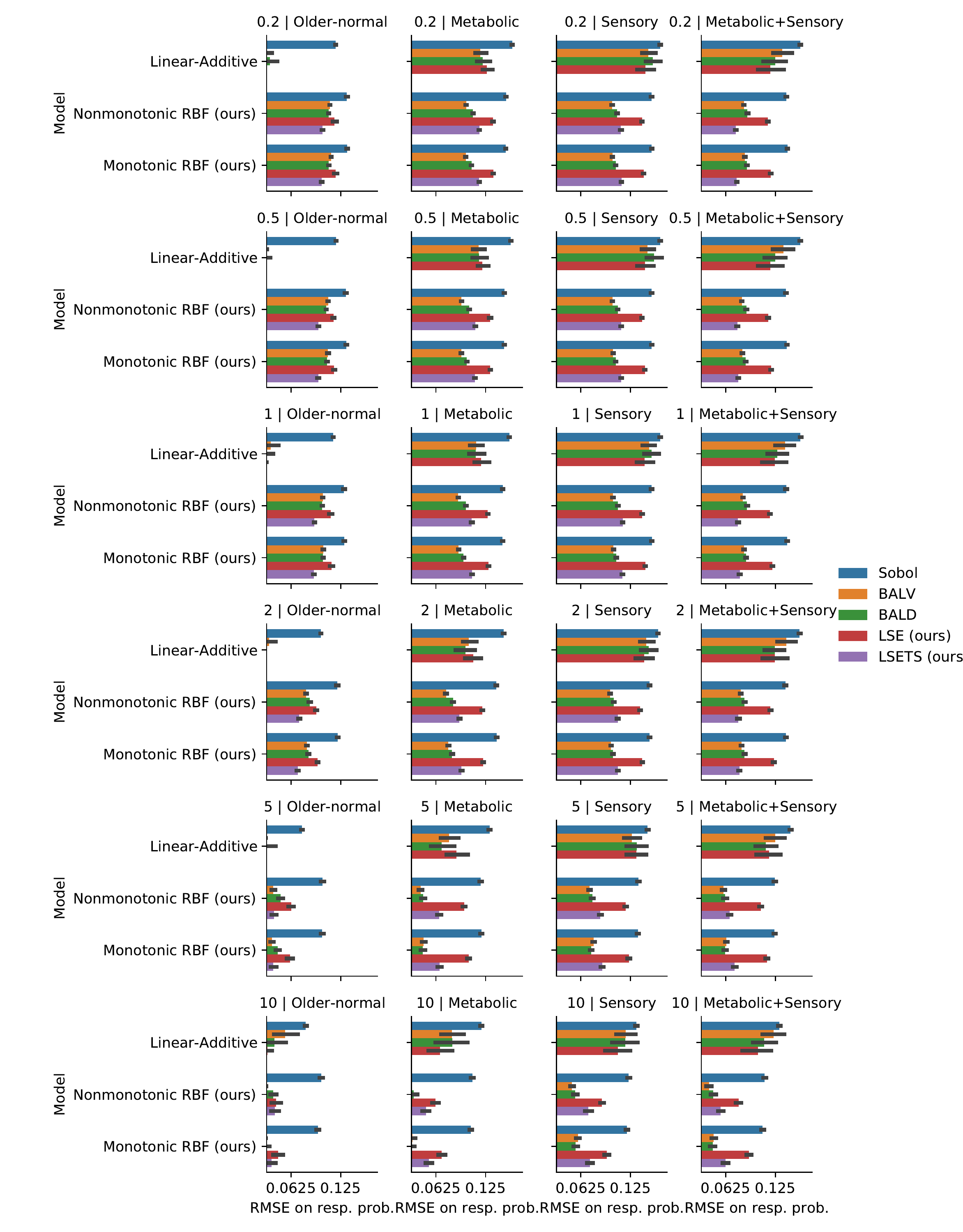}
    \caption{\textbf{Audiometric test function final probability performance}. The overall pattern is heterogenous, but we highlight a few points: first, BALV and BALD are superior to LSE and LSETS when it comes to estimating the full psychometric function, consistent with the latter's focus on estimating the threshold only. Second, the RBF models outperform the linear-additive model, except for the older-normal phenotype. Error bars are a 0.95 confidence interval over simulations. Note the log scale on the x-axis. }
    \label{fig:song-p}
\end{figure*}

\begin{figure*}[!htb]
    \centering
    \includegraphics[height=\textheight]{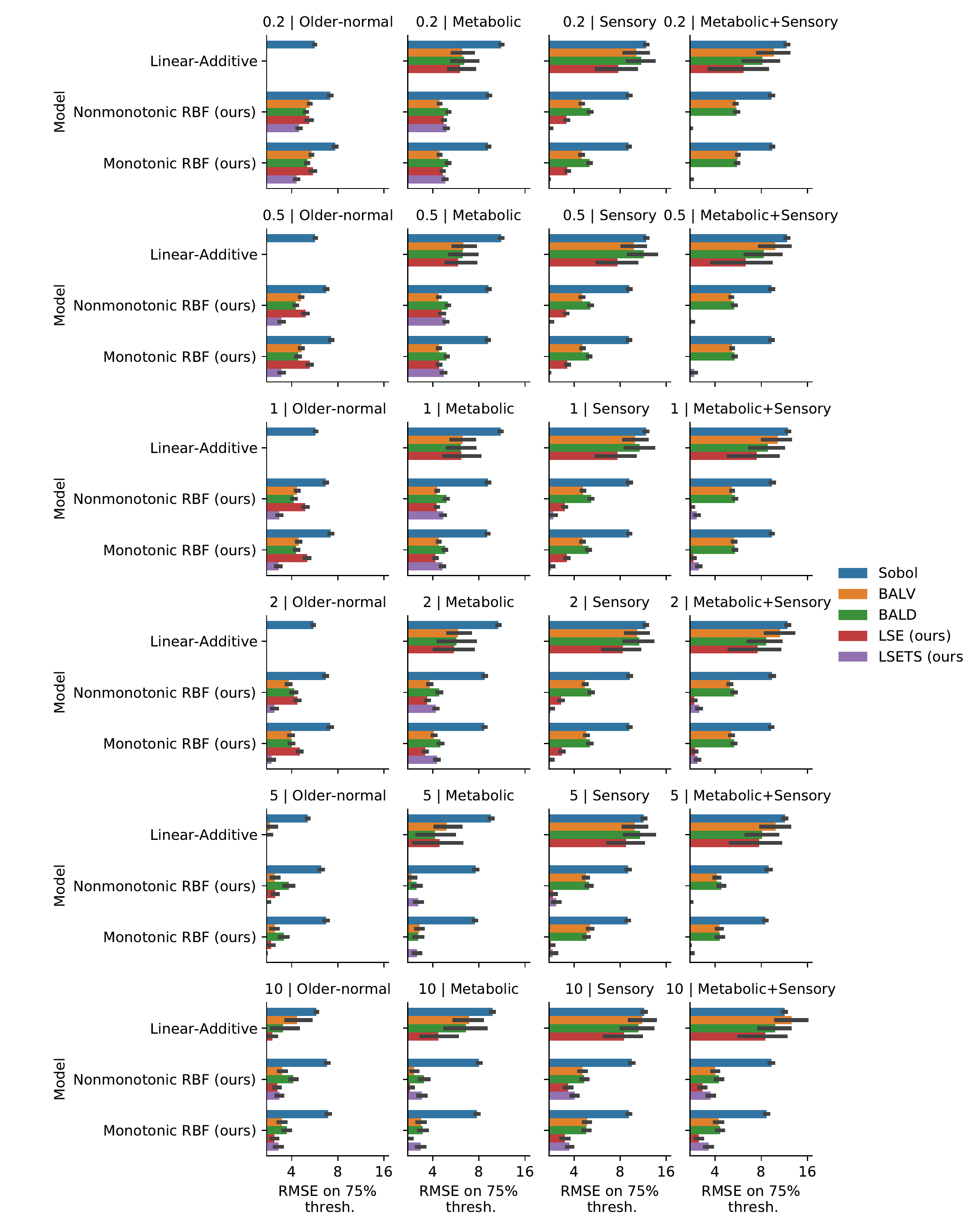}
    \caption{\textbf{Audiometric test function final threshold performance}. The overall pattern is heterogenous, but we highlight a few points: first, the LSE and LSETS objectives consistently perform well for threshold estimation, consistent with their focus on threshold estimation (though sometimes BALD and BALV do well also). Second, the RBF models outperform the linear-additive model, except for the older-normal phenotpye. Errorbars are a 0.95 confidence interval over simulations. Note the log scale on the x-axis}
    \label{fig:song-thresh}
\end{figure*}

Finally, we report the final performance for all individual audiometric test functions in terms of error probability in Fig.~\ref{fig:song-p} and threshold in Fig.~\ref{fig:song-thresh}. In this breakout we see one outlier, the older-normal phenotype, where the linear-additive kernel model outperforms the RBF models (though even here, the LSE objective is superior to BALD and BALV for threshold estimation). We discuss this discrepancy below.

Surprisingly, we did not see a substantial benefit from the monotonic model in our evaluations. While numerically it seems like the monotonic model outperforms the unconstrained RBF model at very small numbers of trials (under about 25), the effect washes out later in the experiment. We suspect this is because for all the test functions we evaluated, the monotonicity property is relatively easy to learn from data: there are large areas of the space where the response probability is either 0 or 1. We suspect that the underperformance of the monotonic model, when it happens, is due to reduced exploration driven by increased confidence, similarly to the linear-additive model. Nonetheless, we think monotonicity is an important addition to the model for theoretical soundness and interpretability by practitioners, as discussed earlier.

\subsection{Results for novel test functions}

\subsubsection{Novel detection test function}
\begin{figure*}[!htb]
    \centering
    \includegraphics[width=\textwidth]{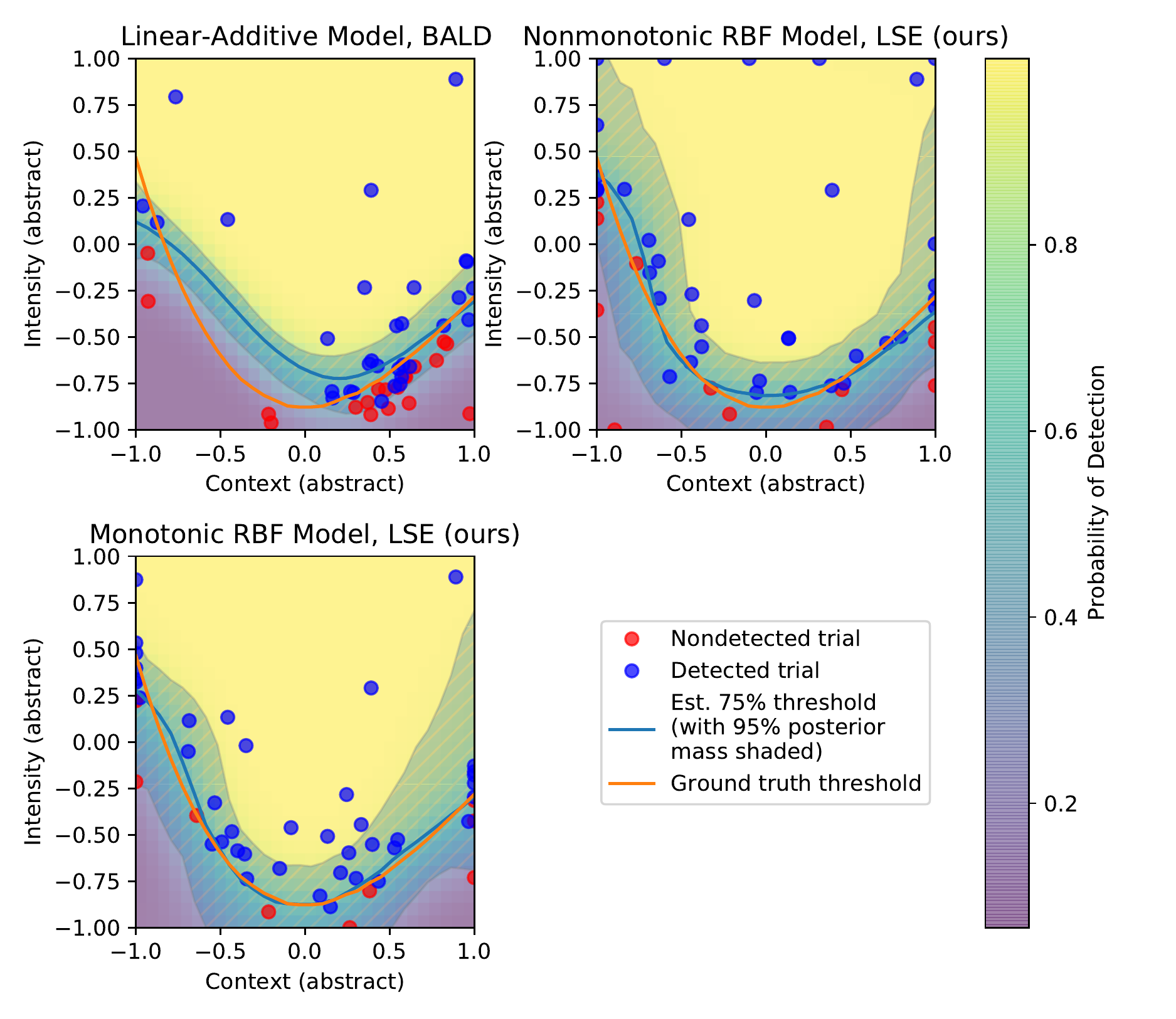}
    \caption{\textbf{Example of samples taken by three representative models on the novel detection test function after 50 trials}. All models begin with the same 5 trials generated from a Sobol sequence and then proceed according to their acquisition function (BALD in the case of the linear-additive model; LSE otherwise). The linear-additive model oversamples the right side of the space, likely due to over-confidence driven by the additive structure. The final estimate does not overlap the true threshold on the left of the plot. Both RBF models with the LSE objective produce good mean threshold estimates. Note that the nonmonotonic model takes multiple samples at the upper and lower boundaries of the space, whereas the monotonic model does not because the monotonicity constraint makes it highly unlikely that the threshold is located there.}
    \label{fig:detection_example}
\end{figure*}

\begin{figure*}[!htb]
    \centering
    \includegraphics[width=\textwidth]{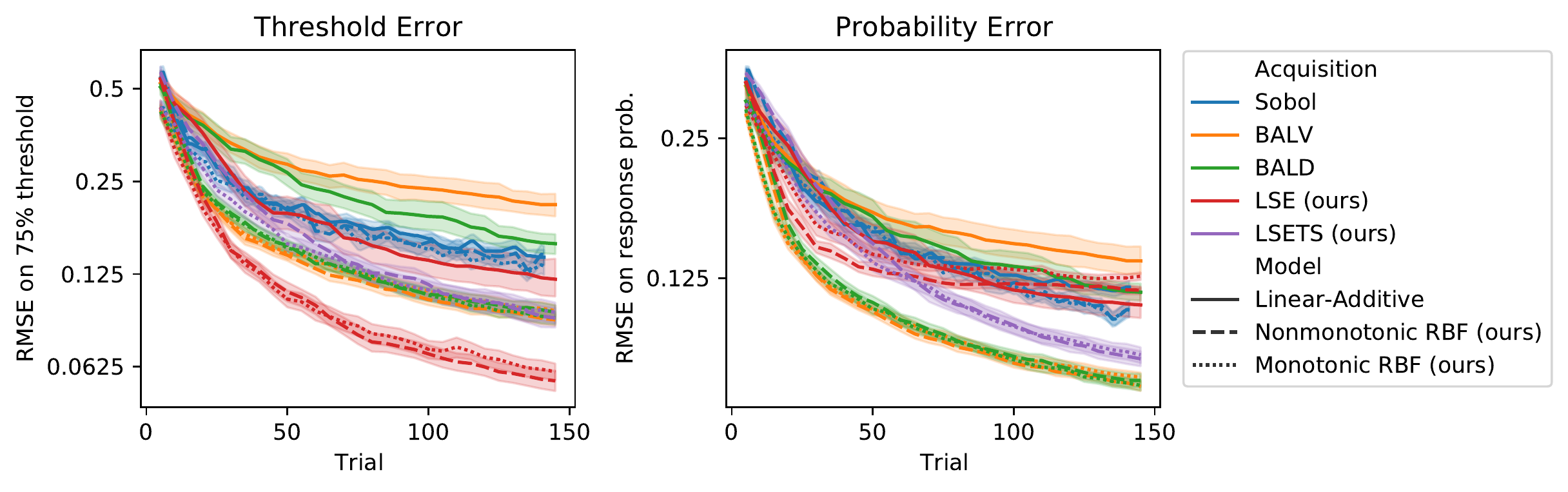}
    \caption{\textbf{Novel detection test function performance}. LSE is competitive in terms of error in both threshold (\emph{left}) and probability (\emph{right}), followed by BALV/BALD and LSETS. For very small numbers of trials, the linear additive model's strong priors allow it to outperform the more flexible models, but with more data the restrictive assumptions hold it back. The monotonic model generally fails to outperform the nonmonotonic one. Shaded intervals are a 0.95 confidence interval over simulations. }
    \label{fig:detection_perf}
\end{figure*}

As with the audiometric test functions, we begin with illustrative examples. A run from the novel detection test function is shown in Fig.~\ref{fig:detection_example}, where we see the same clustered sampling behavior for the linear-additive model seen in the audiometric example. The overall threshold shape appears correct, but the model is expectedly inaccurate in areas where it undersampled. The RBF kernel model performs well regardless of the monotonicity constraint, and while there may appear to be a slight difference in threshold accuracy in the figure, it is not seen in aggregate data. This aggregate data is seen in Fig.~\ref{fig:detection_perf}, where the pattern is very similar to the audiometric setting: the RBF models far outperform the linear-additive model, and LSE outperforms the other objectives for threshold estimation while expectedly falling short in error over the probability surface. In contrast to the audiometric setting, the LSETS acquisition function does not provide a middle ground, patterning with the global objectives in threshold error while failing to match them in probability error.

\subsubsection{Novel discrimination test function}
\begin{figure*}[!htb]
    \centering
    \includegraphics[width=\textwidth]{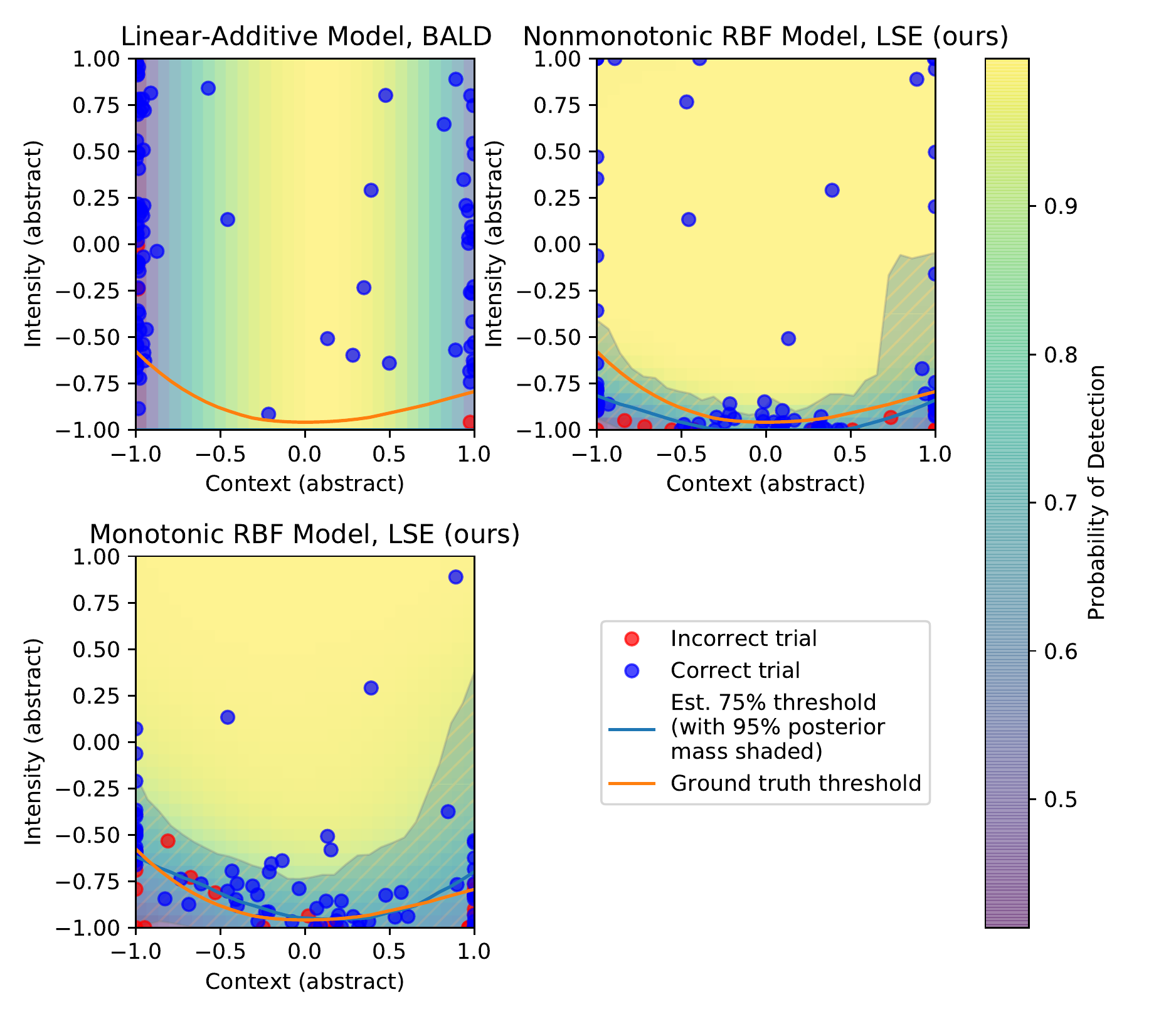}
    \caption{\textbf{Example of samples taken by three representative models on the novel discrimination test function after 100 trials}. All models begin with the same 5 trials generated from a Sobol sequence and then proceed according to their acquisition function (BALD in the case of the linear-additive model; LSE otherwise). We used 100 rather than 50 example trials here because the test function is substantially more difficult than the others. The linear-additive model with BALD fails completely on this task: after observing a few correct trials (consistent with the whole space generating probabilities above 0.5), the model is over-confident over the interior of the space and solely samples at the edges. Both RBF models perform acceptably, though in this case the monotonic model spends fewer trials sampling the upper and side edges (where the threshold is unlikely to be under the monotonicity assumption), and achieves a more accurate mean estimate. This pattern was not consistent on average over replications, however. }
    \label{fig:discrimination_example}
\end{figure*}

A run on the novel discrimination test function is shown for our three representative models in Fig.~\ref{fig:discrimination_example}. This is the hardest test function we explored, in that its domain only covers probabilities above 0.5. Here the linear-additive model with BALD breaks down completely and fails to recover any threshold, spending its entire time sampling stimuli at the edges of the space. In contrast, both RBF models with LSE still estimate reasonable thresholds in this setting. While we see apparent numerical benefits for the monotonic model, these again fail to be borne out in aggregate results, which are shown in Fig.~\ref{fig:detection_perf}. In aggregate, the RBF models we introduce again outperform the linear-additive model, and while the monotonicity constraint provides performance benefits for some acquisition functions, it does not do so for the best-performing ones. When it comes to acquisition, the LSE objective we introduced once more performs best for threshold estimation, though the gap is smaller than in the easier cases. For estimating the full probability surface, LSE surprisingly patterns with the global objectives in this problem.

\begin{figure*}[!htb]
    \centering
    \includegraphics[width=\textwidth]{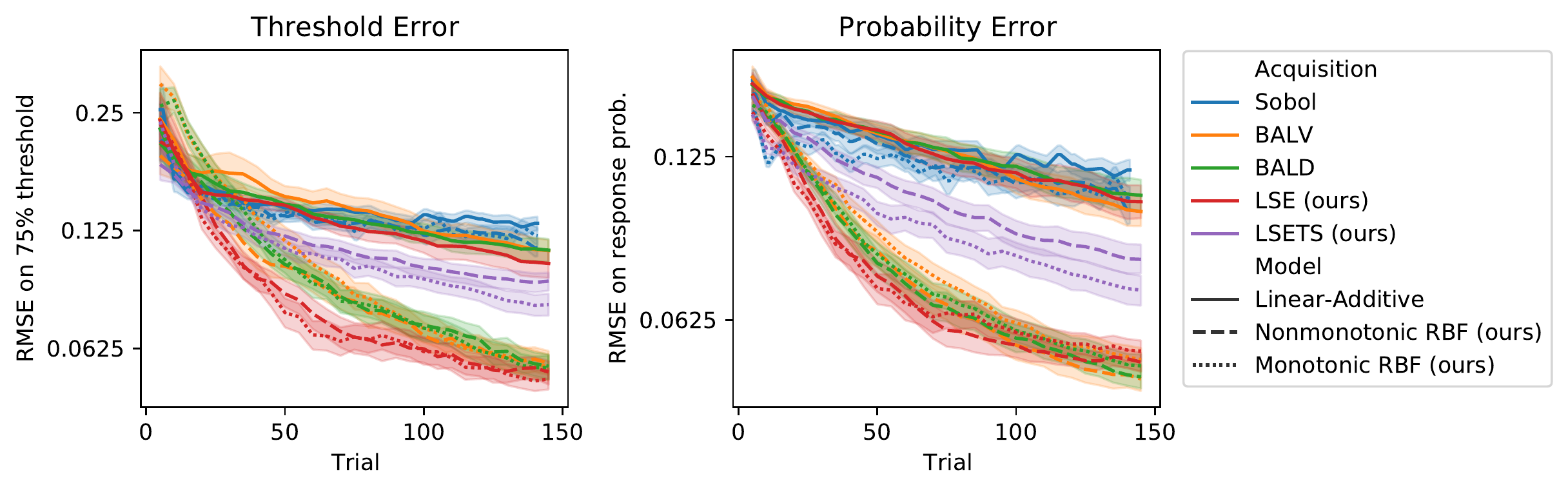}
    \caption{\textbf{Novel discrimination test function performance}. LSE performs best in terms of error in threshold (\emph{left}) whereas the global acquisition objectives BALD and BALV both perform the best in terms of error in probability  (\emph{right}). LSETS shows competitive (but not winning) performance in both. The full-RBF models consistently outperform the linear-additive model, but the monotonic model largely fails to beat the nonmonotonic one. Shaded intervals are a 0.95 confidence interval over simulations. }
    \label{fig:discrimination_perf}
\end{figure*}

\section{Discussion}

Here we remark on a number of interesting patterns seen in our simulation: the curiously poor performance of the linear additive model, the relatively poor performance of the RBF models on the older-normal phenotype audiometric data, and the benefits of LSE over global acquisition functions for threshold estimation.

\subsection{Boundary over-exploration and the poor performance of linear-additive models}

The results above show consistently poor performance for the linear-additive model, even relative to previous reports. As noted above, part of this is driven by the use of variational inference over expectation propagation for inference. In addition, Song and colleagues report excluding the edges of the search space from model evaluation (but not from stimulus generation) ``due to known edge effects of PF estimation." We confirmed (simulations not shown) that evaluating only the interior of the audiometric test functions dramatically improves the performance of the additive-GP model while providing a marginal improvement at best to the full-RBF models, which brings the different model variants much closer to parity (though LSE acquisition still yields best performance).

These results demonstrate a key disadvantage of the linear kernel, in that it trades off between over-exploration and worse estimation of the latent slope. To see why, consider that the posterior covariance of any two points under the linear kernel is the product of their inputs. This means that posterior variance of $f$ will grow quadratically in the intensity dimension. This would generally not be a substantial problem because the probit function will squash very high and very low values to make the posterior variance in probability space small for very high or low probabilities. However, this is only possible if the latent function value can be allowed to go very large or small, which would worsen the estimate of the latent slope (and therefore position of the threshold and the function value itself). Without shrinking the posterior variance at values of the function far from the threshold, variance-sensitive acquisition functions like BALD and BALV will oversample in these areas of the space. Thus, the linear kernel creates an undesirable tradeoff between sampling irrelevant parts of the space and overestimating the slope of the latent psychometric function in the intensity dimension. This problem disappears with the use of the monotonic RBF kernel, which shows less boundary over-exploration than the linear kernel in spite of not using the over-exploration mitigation heuristic of Song and colleagues.

From a practical perspective, while we concur that boundary over-exploration in GPs is a known problem \citep[e.g.][]{Siivola2018}, it is not always practical to extend the search space to extreme stimulus values where thresholds are ignored---for example if these stimuli cannot be physically displayed by the hardware in question. As our objective is the practical usage of these methods across a breadth of psychophysical stimuli, the apparent robustness against boundary over-exploration is an additional benefit of our approach.

\subsection{Understanding the older-normal phenotype performance}

\begin{figure*}[!htb]
    \centering
    \includegraphics[width=\textwidth]{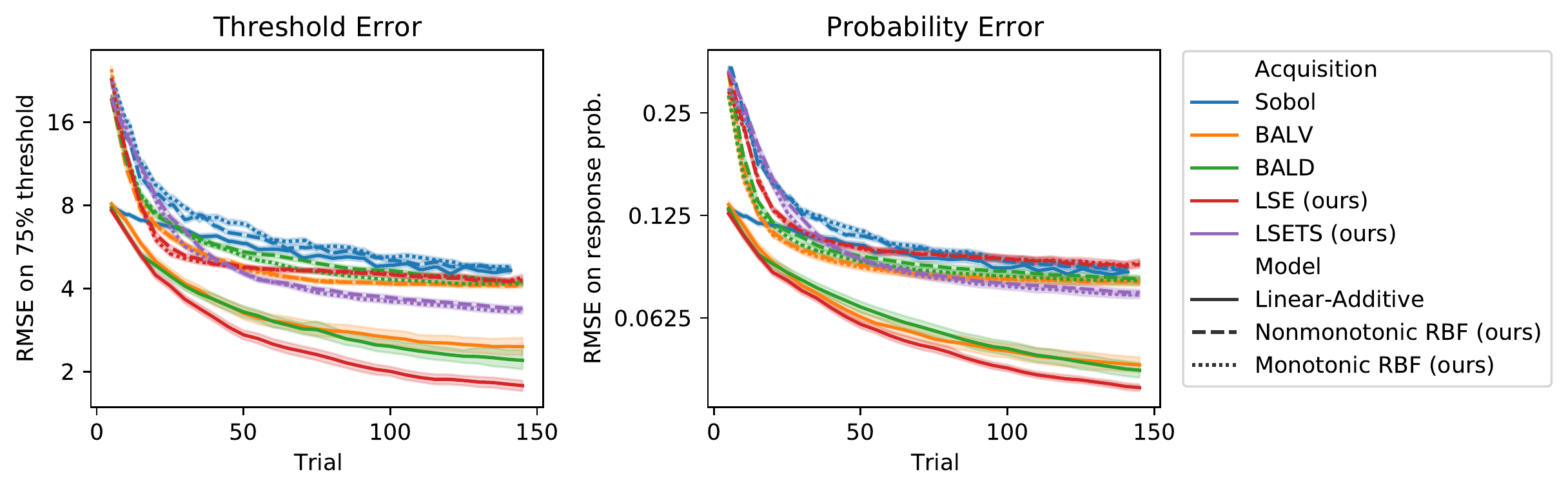}
    \caption{\textbf{Older-normal audiometric test function performance}. Performance is averaged over psychometric spread values. LSE is the best for acquisition in terms of error in both threshold (\emph{left}), and response probability (\emph{right}). The linear-additive model performs far better than the RBF models in this setting, and this performance gap exists from the beginning of sampling, consistent with the idea that this model's prior puts much more probability mass on this test function, which is relatively simpler. Shaded intervals are a 0.95 confidence interval over simulations.}
    \label{fig:song-older-normal}
\end{figure*}

The older-normal audiometric test function is the one case where the linear-additive model consistently performs best. Fig.~\ref{fig:song-older-normal} shows performance for this specific test function, averaged over psychometric spread values. It appears that with very limited data, the linear-additive model's prior is closer to the true threshold than the RBF models can achieve even with a much larger number of trials. Inspecting Fig.~\ref{fig:audiometric-thresholds} shows that in this phenotype the threshold is essentially flat as a function of frequency, and it would appear that the stronger prior of the linear-additive model allows it to capture that property much sooner. We see this as part of a broader pattern where simpler psychometric fields are more easily addressed by simpler models, and more complex ones (such as our novel discimination example) are better tackled with more flexible methods such as ours.

\section{Conclusion}

In this work, we have outlined a new nonlinear generalization of traditional psychophysics theory, where a stimulus-dependent Fechnarian relationship can yield rich nonlinear psychometric transfer functions. This generalized model, expressed as a Gaussian process (GP) with monotonicity information, allows for data-driven estimation of psychometric fields that respect theoretical constraints while remaining more flexible than either classical methods or other methods based on gaussian processes. To support this estimation, we have introduced to the field a new acquisition objective for adaptive psychometric testing targeting perceptual thresholds, the LSE. We have established the benefit of both contributions in extensive simulations, and have provided a software toolbox for application of these methods in real, human-subjects experiments.

\section{Acknowledgements}
We are grateful to Deborah Boehm-Davis, Phil Guan, and Jess Hartcher-O'Brien for helpful comments on the manuscript, and to Eytan Bakshy and Max Balandat for their help and support in applying GPyTorch and BoTorch to this domain.

\section{Open Practices Statement}
AEPsych is available at \url{https://github.com/facebookresearch/aepsych}, alongside the code used to generate this paper and all the results within it.

\clearpage
\printbibliography[]

\end{document}